\documentclass[journal]{IEEEtran}

\usepackage[OT1]{fontenc}
\usepackage[numbers,sort&compress]{natbib}
\usepackage[cmex10]{amsmath}
\usepackage{amssymb}
\usepackage{bm}
\usepackage[dvipdfmx]{graphicx}
\usepackage{color}
\usepackage{subfigure}
\usepackage{mathtools}
\usepackage{multirow}
\usepackage{algorithm,algorithmic}
\usepackage{listings}
\usepackage{comment}
\usepackage{hyperref}
\usepackage{pmat}
\usepackage{tikz}
\usepackage{hyperref}
\usepackage[absolute,overlay]{textpos}

\def\I{\mathbf{I}}
\def\E{\mathbf{E}}
\def\U{\mathbf{U}}

\def\Y{\mathbf{Y}}
\def\e{\mathbf{e}}
\def\Yh{\hat{\mathbf{Y}}}

\def\H{\mathbf{H}}

\def\S{\mathbf{S}}

\def\V{\mathbf{V}}

\def\X{\mathbf{X}}
\def\Xh{\hat{\mathbf{X}}}

\def\A{\mathbf{A}}
\def\M{\mathbf{M}}
\def\P{\mathbf{P}}

\newcommand{\argmin}{\mathop{\rm arg~min}\limits}
\newcommand{\maximize}{\mathop{\rm maximize}\limits}
\newcommand{\minimize}{\mathop{\rm minimize}\limits}

\RequirePackage[normalem]{ulem} 
\RequirePackage{color}\definecolor{RED}{rgb}{1,0,0}\definecolor{BLUE}{rgb}{0,0,1} 

\makeatletter
\def\ps@IEEEtitlepagestyle{%
  \def\@oddfoot{\mycopyrightnotice}%
  \def\@oddhead{\hbox{}\@IEEEheaderstyle\leftmark\hfil\thepage}\relax
  \def\@evenhead{\@IEEEheaderstyle\thepage\hfil\leftmark\hbox{}}\relax
  \def\@evenfoot{}%
}
\def\mycopyrightnotice{%
  \begin{minipage}{\textwidth}
  \centering \scriptsize
\textcopyright 2023 IEEE. Personal use of this material is permitted.
  Permission from IEEE must be obtained for all other uses, in any current or future
  media, including reprinting/republishing this material for advertising or promotional
  purposes, creating new collective works, for resale or redistribution to servers or
  lists, or reuse of any copyrighted component of this work in other works.
  DOI: \href{https://doi.org/10.1109/TIFS.2023.3277255}{10.1109/TIFS.2023.3277255}
  \end{minipage}
}
\makeatother
\begin{document}

\title{Noncoherent Massive MIMO with Embedded One-Way Function Physical Layer Security}

\author{Yuma~Katsuki,~\IEEEmembership{Member,~IEEE},
Giuseppe~Thadeu~Freitas~de~Abreu,~\IEEEmembership{Senior Member,~IEEE},\\
Koji~Ishibashi,~\IEEEmembership{Senior Member,~IEEE}, and Naoki~Ishikawa,~\IEEEmembership{Senior Member,~IEEE}.\thanks{Yuma~Katsuki and Naoki Ishikawa are with the Faculty of Engineering, Yokohama National University, 240-8501 Kanagawa, Japan (e-mail: ishikawa-naoki-fr@ynu.ac.jp). Giuseppe Thadeu Freitas de Abreu is with the School of Computer Science and Engineering, Constructor University, 28759 Bremen, Germany (e-mail: gabreu@constructor.university). Koji~Ishibashi is with the Advanced Wireless and Communication Research Center, The University of Electro-Communications, 182-8585 Tokyo, Japan (e-mail: koji@ieee.org). This work was supported in part by the Japan Science and Technology Agency, Strategic International Collaborative Research Program (JST SICORP), Japan, under Grant JPMJSC20C1. Part of this paper was presented at IEEE VTC2021-Fall \cite{katsuki_optimal_2021}.}}
\maketitle

\TPshowboxesfalse
\begin{textblock*}{\textwidth}(45pt,10pt)
\footnotesize
\centering
Accepted for publication in IEEE Transactions on Information Forensics and Security. This is the author's version which has not been fully edited and content may change prior to final publication. Citation information: DOI 10.1109/TIFS.2023.3277255
\end{textblock*}

\begin{abstract}
We propose a novel physical layer security scheme that exploits an optimization method as a one-way function. The proposed scheme builds on nonsquare differential multiple-input multiple-output (MIMO), which is capable of noncoherent detection even in massive MIMO scenarios and thus resilient against risky pilot insertion and pilot contamination attacks. In contrast to conventional nonsquare differential MIMO schemes, which require space-time projection matrices designed via highly complex, discrete, and combinatorial optimization, the proposed scheme utilizes projection matrices constructed via low-complexity continuous optimization designed to maximize the coding gain of the system. Furthermore, using a secret key generated from the true randomness nature of the wireless channel as an initial value, the proposed continuous optimization-based projection matrix construction method becomes a one-way function, making the proposed scheme a physical layer secure differential MIMO system. An attack algorithm to challenge the proposed scheme is also devised, which demonstrates that the security level achieved improves as the number of transmit antennas increases, even in an environment where the eavesdropper can perfectly estimate channel coefficients and experience asymptotically large signal-to-noise ratios.
\end{abstract}


\IEEEpeerreviewmaketitle

\section{Introduction}
\IEEEPARstart{P}{hysical} layer security (PLS) is an increasingly important research topic in wireless communications towards the post-quantum era.
Considering that radio waves can propagate over long distances, and the ever-increasing demand for wireless-enabled, data-hungry and sensitive applications, an enormous amount of confidential information is exposed over the air.
The current approach to secure such information is fundamentally reliant on encryption methods introduced at higher layers, which
however can be vulnerable to emerging (in particular quantum) technologies.
The classic asymmetric Rivest, Shamir, and Adleman (RSA) encryption method \cite{rivest_method_1978}, for instance, relies on the complexity of prime factorization, which is threatened by Shor's algorithm \cite{shor_algorithms_1994} in case a large-scale fault-tolerant quantum computer is employed.

It is known that noise has a deleterious effect on the accuracy of quantum circuits \cite{fujii_noise_2016}, which opens a gap between the theoretical capabilities of Shor's algorithm and its practical performance of when implemented in real-world quantum computers \cite{amico_experimental_2019}.
While this challenge can win traditional cryptography some time, it is expected that such engineering hurdles will be gradually overcome \cite{BallNature2021}.

In light of the above, the main idea of PLS is to exploit physical features of wireless channels \cite{dean_physical-layer_2017,allen_secure_2014,althunibat_physical-layer_2017,okamoto_chaos_2012,ishikawa_articially_2021}, to design quantum resilient, and eventually quantum-proof (a.k.a post-quantum) security solutions.
To cite a few examples, in \cite{dean_physical-layer_2017}, the randomness of extracted channel state information (CSI) in massive MIMO systems is used to construct linear precoding matrices that provide post-quantum security via large-scale beamforming and large constellation sizes.
Similarly, in \cite{allen_secure_2014}, security is incorporated into STBC schemes by translating random received signal strength indicators into phase rotations of transmit symbols, while in \cite{althunibat_physical-layer_2017}, the reciprocal phase of the channel between Alice and Bob is used as a seed for the legitimate pair to select the modulation they employed.
In turn, in the chaos-based approaches of \cite{okamoto_chaos_2012} and \cite{ishikawa_articially_2021}, a secret key is projected onto a space-time codeword and onto a time-varying unitary matrix, respectively.

All of these schemes have two security-enhancing components incorporated, namely, the possibility of frequent updates of channel-based keys, and the one-way projection which makes it difficult for an eavesdropper to estimate the key from the original transmit signals.
The major drawback of the approaches is, however, the underlying assumption of perfect CSI (PCSI) knowledge \cite{huang_fast_2013, zeng_physical_2015} by Alice and Bob, which implies frequent exchanges of pilot signals that in turn can be exploited by eavesdroppers to obtain precise CSI \cite{zhang_design_2017, hamamreh_ofdm-subcarrier_2017}.

In contrast, differential space-time modulation \cite{hughes_differential_2000} does not require CSI estimation, such that the integration of the latter with the aforementioned approaches can be leveraged to reduce the risk of eavesdropping attacks in one-way function-based security schemes.
However, the seminal approach \cite{hughes_differential_2000} requires an $M \times M$ square unitary matrix to modulate $M$ symbols transmitted by $M$ transmit antennas, which degrades the effective transmission rate by a factor of $M$.
Fortunately, extended differential space-time modulation schemes have recently emerged \cite{ishikawa_rectangular_2017, ishikawa_differential_2018, ishikawa_differential-detection_2019, xiao_differentially-encoded_2020}, which rely on $M \times T$ nonsquare space-time projection matrices, with $T < M$, to modulate $M$ symbols, such that higher transmission rates can be achieved, with particular relevance in massive MIMO scenarios.
In particular, it is shown in \cite{ishikawa_differential_2018} that the nonsquare counterpart of the diagonal unitary coding (DUC) \cite{hochwald_differential_2000} achieves competitive performance, although it requires the solution of a highly-complex discrete optimization problem to be implemented, especially as the number of transmit antennas increases.

Against this background, we conceive a low-complexity but optimal optimization method for nonsquare differential massive MIMO coding.
Based on this method, we then propose a PLS scheme that utilizes the optimization method as a one-way function, thus enabling asymmetric MIMO-based encryption exploiting features of the wireless channel.
The contributions of the article can be summarized as follows.
\begin{itemize}
\item \textbf{We conceive a low-complexity but optimal construction method} for the square-to-nonsquare projection matrix. The conventional DUC-aided nonsquare coding requires a high complexity discrete optimization for unitary matrices. The corresponding time complexity increases exponentially with the number of transmit antennas and transmission rate, resulting in a challenge in open-loop massive MIMO scenarios. We generalize the square-to-nonsquare projection to enable continuous optimization. Then, we derive an objective function that is low-complexity despite achieving equivalent performance to the conventional counterpart, and demonstrate that the proposed scheme is optimal in terms of the coding gain and bit error rate.
\item \textbf{We propose a novel concept that utilizes an optimization method as a one-way function}. To the best of the authors' knowledge, this is the first attempt of that kind in the context of wireless PLS. We compare representative optimization methods in terms of time complexity and optimality, and identify a good optimization method that can reduce the optimization delay while maintaining a good performance. In terms of the information leakage and the secrecy rate, we show that the proposed scheme with a large number of transmit antennas is capable of achieving high security even under extreme conditions where the eavesdropper can estimate perfect channel coefficients and can ignore noise.
\end{itemize}

The remainder of the article is organized as follows. The system model assumed throughout the article is defined in Section \ref{sec:sys}. The conventional nonsquare differential coding scheme is briefly reviewed in Section \ref{sec:conv:n}. In Section \ref{sys:pnon}, the new optimization method to design square-to-nonsquare projection matrices is proposed and shown to be both optimal and of low complexity. In Section \ref{sec:ppro}, the application of the new optimization method as PLS protocol is described, along with an attack algorithm to probe the efficacy of the proposed scheme. The security performance of the scheme is analyzed in Section \ref{sec:comp}, and finally, concluding remarks are offered in Section \ref{sec:conc}.

\section{System Model\label{sec:sys}}

We assume that a legitimate transmitter Alice, equipped with $M$ transmit antennas, communicates with a legitimate receiver Bob, equipped with $N$ receive antennas, in the presence of eavesdropper Eve, also equipped with $N$ receive antennas, but with unlimited computing capabilities, such that the received signal at Bob is given by

\begin{equation}
\Y(i) = \H(i)\S(i) + \V(i) \in \mathbb{C}^{N \times T},
\label{COH:Bob:blockmodel}
\end{equation}
where $i$ denotes a transmission index, $\H(i) \in \mathbb{C}^{N \times M}$ denotes the independent and identically distributed (i.i.d.) small-scale Rayleigh fading channel matrix such that $h_{n,m}\sim\mathcal{CN}(0,1)$, $\S(i) \in \mathbb{C}^{M \times T}$ denotes a space-time codeword, and $\V(i) \in \mathbb{C}^{N \times T}$ denotes the i.i.d complex additive white Gaussian noise (AWGN) matrix such that $v_{n,t}\sim\mathcal{CN}(0,\sigma_v^2)$, with $1 \leq n \leq N$, $1 \leq m \leq M$, $1 \leq t \leq T$ and the per-symbol signal-to-noise ratio (SNR) given $1/\sigma_v^2$.

Similarly, the received signal at Eve is given by
%
\begin{equation}
\Y_{\mathrm{Eve}}(i) = \H_{\mathrm{Eve}}(i)\S(i) + \V_{\mathrm{Eve}}(i) \in \mathbb{C}^{N \times T},
\label{COH:Eve:blockmodel}
\end{equation}
where $\H_{\mathrm{Eve}}(i)$ and $\V_{\mathrm{Eve}}(i)$ are analogous of their counterparts at Bob.

We remark that in the security performance evaluation, it will assume that Eve not only has perfect knowledge of $\H_{\mathrm{Eve}}(i)$ but also, in an ideal case, may be subject to an infinite SNR, $i.e.$, $\V_{\mathrm{Eve}}(i)=\mathbf{0}$.

\section{Conventional Nonsquare Differential Coding}
\label{sec:conv:n}

In this section, we briefly review conventional nonsquare-matrix-based differential space-time coding (N-DSTC) schemes\footnote{Notice that noncoherent massive MIMO schemes \cite{chen-hu_non-coherent_2020, xie_non-coherent_2020} employ simple modulation techniques such as PAM and PSK, and thus differ from the N-DSTC scheme which uses a unitary matrix and its nonsquare projection.}, including specific construction methods for space-time codewords, encoding and decoding procedures.
The design of the projection matrices $\E_i$ is an integral part of the design of a given N-DSTC scheme, and can be considered a form of physical encryption applied over the space-time codewords.
The design of these matrices will be revisited in detail later, and is at the core of our contribution.

\subsection{Space-Time Modulation: DUC and ADSM Methods}
\label{subsec:stm}

N-DSTC \cite{ishikawa_differential_2018} was proposed as a method to circumvent the requirement for perfect CSI estimation of classic coherent systems and to increase the transmission rate of square-matrix based DSTC \cite{hughes_differential_2000,hochwald_differential_2000,rajashekar_algebraic_2017,tarokh_differential_2000}, which become particularly challenging in high-speed mobile and massive MIMO scenarios.

In N-DSTC schemes, $B$ bits of information are mapped into a data matrix $\X(i)\in\mathbb{C}^{M \times M}$ selected from a codebook of $2^B$ matrices.
Among others, two space-time codebook construction methods are representative: a) the DUC method of \cite{hochwald_differential_2000}, and b) the algebraic differential spatial modulation (ADSM) scheme of \cite{rajashekar_algebraic_2017}.

In the DUC method \cite{hochwald_differential_2000}, the $B$ input bits are mapped into the integers $b=0,1, \cdots, 2^B-1$, and the corresponding unitary matrix is generated by
\begin{equation}
\X(i) = \mathrm{diag}\bigg[\mathrm{exp}\Big(j\frac{2\pi b}{2^B}u_1\Big), \cdots, \mathrm{exp}\Big(j\frac{2\pi b}{2^B}u_M\Big)\bigg],
\label{COH:DUC:datamatrix}
\end{equation}
where $j$ denotes the elementary imaginary number.

In this scheme, diversity is maximized by ensuring that the $M$ factors $u_m$ are designed to maximize the diversity product
%
\begin{equation}
P_{\mathrm{max}}=\min_{b\in\{1, \cdots, 2^B-1\}}\bigg|\prod_{m=1}^M \mathrm{sin}\Big(\frac{\pi b u_m}{2^B}\Big)\bigg|^\frac{1}{M}\!\!\!\!\!,
\label{COH:DUC:fanc}
\end{equation}
while satisfying the condition $0 < u_1 \leq \cdots u_M \leq2^B/2 \in \mathbb{Z}$.

Notice that the search space size of problem \eqref{COH:DUC:fanc} can be calculated as ${2^{B-1}+M-1 \choose M}$, such that for $M = 32,64,128,$ and $256$ transmit antennas, we have ${3M-1 \choose M}$ candidates to map $B=\mathrm{log}_{2}(M)+2$ bits, which corresponds to $2\times10^{25}, 5\times10^{51}, 4\times10^{104},$ and $4\times10^{210}$ candidates, respectively, making problem \eqref{COH:DUC:fanc} highly complex.

In turn, in the ADSM method \cite{rajashekar_algebraic_2017},
the $B$ input bits are partitioned into two sequences of length $B_1=\log_2(M)$ and $B_2=\log_2(L)$ bits, respectively, with the first $B_1$ bits used to select a specific dispersion matrix (DM) out of a codebook of DM matrices $\A_m$ given by
%
\begin{equation}
\mathcal{A}\triangleq\{\A_1, \cdots, \A_M\} = \{\I_M, \M, \M^2, \cdots, \M^{M-1}\},
\label{COH:eq:ADSM:batrix}
\end{equation}
%
\begin{equation}
\M \triangleq
\begin{pmat}[{|}]
\mathbf{0}_{1\times (M-1)} & e^{j 2\pi/L} \cr
\-
\I_{M-1} & \mathbf{0}_{(M-1)\times 1} \cr
\end{pmat}\in \mathbb{C}^{M \times M},
\label{COH:eq:permutation:matrix}
\end{equation}
which implies that $\mathcal{A}$ is a set of unitary matrices.

At each $i$-th transmission block, a given DM $\A_m(i)$ is selected to encode the first $B_1$ bits, while the remaining $B_2$ bits are mapped to a symbol $s(i)\in\mathcal{S}$, with $\mathcal{S}$ denoting the $L$-PSK constellation, such that the following data matrix is generated
\begin{equation}
\X(i) = s(i)\A_m(i).
\label{COH:ADSM:datamatrix}
\end{equation}

\subsection{Nonsquare Differential Encoding}
\label{subsec:nonsq_encoding}

To elaborate, in the N-DSTC scheme, at the beginning of a transmission stream, the set of projection matrices $\{\E_1, \cdots, \E_{M/T}\}$ are transmitted.
During this initial stage, the received symbol is expressed as
\begin{equation}
\Y(i)=\H(i)\E_i+\V(i),
\label{COH:transmit:pilot}
\end{equation}
for $1 \leq i \leq M/T$. Notice that this requires $M$ timeslots.

After the projection matrices are communicated to the receiver, and assuming a  transmission frame of $W$ blocks, the subsequent data matrices $\X(i)$ are differentially encoded as
\begin{align}
\S(i) = \left\{ \begin{array}{ll}
\I_M & \text{if } i=M/T, \text{or}\\
\S(i-1)\X(i) & \text{if } i>M/T,
\end{array} \right.
\label{COH:differencial}
\end{align}
mapped onto $M \times T$ nonsquare matrices via multiplication with the projection matrix $\E_1$, and transmitted such that the corresponding received signal becomes
\begin{equation}
\Y(i)=\H(i)\S(i)\E_1+\V(i),
\label{COH:transmit:data}
\end{equation}
for $M/T < i \leq W$.

The ratio between the number of transmit antennas $M$ and the frame length $W$, $i.e., \eta \triangleq M/W$, is referred to as the reference insertion ratio.
For the sake of performance assessment, we will set $\eta$ to 5\%, such that $W=20M$, but we remark that the performance of N-DSTC schemes remain robust in high-speed mobile scenarios \cite{ishikawa_differential-detection_2019} even with $\eta= 1\%$ and $\eta= 0.1\%$, which yield frames of length $W=100M$ and $W=1000M$, respectively.

\subsection{Algebraic Construction of Nonsquare Projection Basis}
\label{subsec:basis}

The N-DSTC scheme relies on a basis of projection matrices $\E_i\in \mathbb{C}^{M \times T}$, constructed to satisfy the constraints \cite{ishikawa_differential_2018}:
\begin{subequations}
\begin{equation}
\text{Power constraint}:\; \|\E_i\|^2_F=T,
\label{COH:power:constration}\\
\end{equation}
\begin{equation}
\text{Normality}:\; \E^H_i\E_i=\I_T,
\label{COH:normality}\\
\end{equation}
\begin{equation}
\text{Orthogonality}:\;\E^H_i\E_{i'}=\mathbf{0}_T,
\label{COH:orthogonality}\\
\end{equation}
\begin{equation}
\text{Reconstructibility}:\; \sum_{i=1}^{M/T} \E_i\E^H_i=\I_M,
\label{COH:reconstructibility}
\end{equation}
\end{subequations}
for all $1\leq i\neq i' \leq M/T$.

A strategy to obtain the projection matrix basis is to partition a unitary matrix $\U_M \in \mathbb{C}^{M \times M}$, such that $\E_i$ is given by the $(i-1)T$-th to the $iT$-th columns of $\U_M$, that is
\begin{equation}
\U_M =
\begin{pmat}[{|||}]
\E_1 & \E_2 & \cdots  & \E_{M/T}\cr
\end{pmat}.
\label{eq:unitary_to_E}
\end{equation}

In turn, in \cite{ishikawa_differential_2018} it was proposed to construct the unitary matrix $\U_M$ for a system with $M$ transmit antennas, and given a number $N_b$ of nonzero components in each column, as follows
\begin{equation}
\U_M \triangleq \mathrm{bdiag}\underbrace{[\mathbf{W}, \cdots, \mathbf{W}]}_{M/N_b~\text{repetition}}\in \mathbb{C}^{M \times M},
\label{COH:hybrid:basis}
\end{equation}
with $\mathbf{W}$ denoting the discrete Fourier transform matrix

\begin{equation}
	\mathbf{W} = \frac{1}{\sqrt{N_b}}\!\left[\begin{array}{cccc}
	1 & 1 & \cdots & 1\\
	1 & \omega & \cdots & \omega^{N_b-1}\\
	1 & \omega^2 & \cdots & \omega^{2(N_b-1)}\\
	\vdots & \vdots & \ddots & \vdots\\
	1 & \omega^{N_b-1} & \cdots  &\omega^{(N_b-1)(N_b-1)}
	\end{array}\right]\! \in \mathbb{C}^{N_b \times N_b},
	\label{COH:DFT:matrix}
\end{equation}
where $\omega\triangleq\mathrm{exp}(-2\pi j/N_b)$.

Notice that $\E_1$ becomes more sparse as $N_b\to 1$, and denser as $N_b\to M$.
Also, since $M/N_b$ has to be an integer, we shall set $N_b = 1, 2^1, 2^2, \cdots,  2^{\mathrm{log_2}(M)} = M$ in this paper.

\subsection{Noncoherent ML Detection}
\label{subsec:nonsq_detection}
In light of the design described above, a simple (if suboptimal) noncoherent N-DSTC receiver is \cite{ishikawa_differential_2018}
\begin{equation}
\Xh(i)=\argmin_{\X} ||\Y(i)-\Yh(i-1)\X\E_1||^2_F,
\label{COH:nonsq:detection}
\end{equation}
where the key component $\Yh(i)$ is defined by
\begin{equation}
\Yh(i) = \left\{\!\!
\begin{array}{ll}
\sum_{k=1}^{M/T} \Y(k)\E^H_k & \text{for } i=M/T, \\[1ex]
\Y(i)\E^{(1-\alpha)}+\Yh(i-1)\Xh(i)\E^{(\alpha)} & \text{for } i>M/T,
\end{array} \right.
\label{COH:Y:hat}
\end{equation}
while the constant matrices $\E^{(\alpha)}$ and $\E^{(1-\alpha)}$ are defined as
\begin{subequations}
\label{COH:constant:matrix}
\begin{eqnarray}
&\E^{(\alpha)} \triangleq\alpha(i) \E_1 \E_1^H + \sum_{k=2}^{M/T} \E_k \E_k^H,& \label{COH:constant:matrix1} \\
&\E^{(1-\alpha)} \triangleq(1-\alpha(i))\E_1^H.&
\label{COH:constant:matrix2}
\end{eqnarray}
\end{subequations}

In equations \eqref{COH:constant:matrix1} and \eqref{COH:constant:matrix2}, $\alpha(i)$ is a forgetting factor that determines the ratio of $\Y(i)$ and $\Yh(i-1)$ added together, which can be updated via the closed-form expression  \cite{ishikawa_differential-detection_2019}
\begin{equation}
\alpha(i)=\min(\max(\alpha_v(i),0.01),0.99),
\label{COH:forgetting:factor}
\end{equation}
where
\begin{equation}
\alpha_v(i)\triangleq\frac{N \cdot T \cdot \sigma_v^2}{||\mathbf{D}(i)||^2_F},
\label{COH:forgetting:factor:v}
\end{equation}
with the adaptive forgetting factor $\alpha_v(i)$ changed depending on the SNR and the difference between $\H(i)$ and $\H(i-1)$ \cite{ishikawa_differential-detection_2019}, and $\mathbf{D}(i)\triangleq\Y(i)-\Yh(i-1)\Xh(i)\E_1$.

\section{Proposed Nonsquare Coding}
\label{sys:pnon}

Despite various interesting features, there are a few aspects of the conventional N-DSTC schemes that can be improved.
One of such aspects is that the projection matrix $\E_1$ is not designed specifically for the mapping approach utilized, $e.g.$ DUC or ADSM, and another is the fact that constructing the matrices $\E_2, \cdots, \E_{M/T}$ satisfying the constraints \eqref{COH:power:constration} to \eqref{COH:reconstructibility} can become cumbersome in massive MIMO settings.

In this section, we therefore improve upon the conventional N-DSTC schemes in two ways.
Firstly, an optimal method to construct $\E_1$ in which its nonzero elements are generalized to complex numbers is proposed.
The minimum conditions to achieve the performance upper bound are also described.
Secondly, a novel N-DSTC scheme that relies only on $\E_1$ is designed, which eliminates the need of constructing the matrices $\E_2, \cdots, \E_{M/T}$, reducing the overall complexity of the scheme.

Notice that these two improvements make room for the contribution introduced thereafter, by enabling the construction of $\E_1$ to be revisited in the context of one-way function designs for asymmetric PLS systems.

\subsection{Projection Matrix Design via Continuous Optimization}
\label{subsec:prop:basis}

For starters, let us analyze the performance of N-DSTC when $\E_1$ satisfies only the transmit power constraint of equation \eqref{COH:power:constration}, and identify the minimum condition under which the maximum performance is achieved.
To that end, first consider the simplest when $(N_b, T) = (M, 1)$.
In this case, $\E_1 \in \mathbb{C}^{M \times 1}$ is a vector, which for the sake of clarity is denoted $\e_1 \in \mathbb{C}^{M \times 1}$.
When $\e_1$ satisfies the power constraint and the power is equally distributed among transmit antennas, $\e_1$ can be expressed as
\begin{equation}
\e_1=[e_1\ e_2\ \cdots\ e_M]^T /\sqrt{M} \in \mathbb{C}^{M \times 1},
\label{COH:propose:basis}
\end{equation}
where each $e_m=e^{j\theta_m}$, with $1 \leq m \leq M$ and $0 \leq \theta_m <2\pi$, lies on the unit-radius circle on the complex plane.

A good performance metric to evaluate the projection vector $\e_1$ is the coding gain defined as the minimum, among all pairs,  Euclidean distance between projected data matrices $\X_p \e_1$ and $\X_q \e_1$, that is \cite{hanzo_near-capacity_2009}
\begin{align}
g(\e_1)&\triangleq\min_{\X_p \neq \X_q\in \mathcal{X}} \left((\X_p\e_1-\X_q\e_1)^H(\X_p\e_1-\X_q\e_1) \right)^\frac{1}{N}\nonumber\\
& = \min_{\X_p \neq \X_q\in \mathcal{X}} \left(\e_1^H(\X_p-\X_q)^H(\X_p-\X_q)\e_1 \right)^\frac{1}{N}
\label{COH:coding:gain}
\end{align}
where $\mathcal{X}\triangleq \{\X_1, \cdots, \X_{2^B}\} \in \mathbb{C}^{M \times M}$ is the codebook of all ${2^B}$ possible data matrices $\X$.

The above implies that the optimal projection vector $\mathbf{\e_1}$ is the solution of the problem
\begin{equation}
\begin{aligned}
\maximize_{\mathbf{\e_1}\in \mathbb{C}^{M \times 1}} \quad & g(\e_1) \\
\textrm{s.t.} \quad & |e_m|^2=1, \forall\; m=1,2,\cdots,M,
\end{aligned}
\label{eq:maxge1}
\end{equation}
which is highly complex, as it requires the evaluation of the Euclidean distances of all pairs of $M\times M$ matrices in the set
$\mathcal{X}$, of cardinality $2^B$, at the cost of $\mathcal{O}(M^3)$ per pair.

Thanks to the fact that DUC codewords are diagonal and already optimized via discrete combinatorial optimization \cite{hochwald_differential_2000}, the coding gain defined in equation \eqref{COH:coding:gain} does not depend on $\e_1$ in the particular case of DUC mapping.
More generally, and in the case of ADSM, however, the coding gain of equation \eqref{COH:coding:gain} varies with $\e_1$, requiring a solution of the highly complex problem \eqref{eq:maxge1}.
In order to avoid the associated complexity, we first show in Appendix \ref{app:coding:gain},  that equation \eqref{COH:coding:gain} can be simplified to
\begin{equation}
g(\e_1)=\min(g_1(\e_1), g_2(\e_1)),
\label{COH:min:coding:gain}
\end{equation}
with the functions $g_1(\e_1)$ and $g_2(\e_1)$ respectively given by
\begin{equation}
g_1(\e_1)=\min_{s_1, s_2 \in \mathcal{S}}
\left|s_1-s_2\right|^\frac{2}{N},\label{COH:propose:G1}
\end{equation}
\begin{equation}
g_2(\e_1)=\!\!\!\!\!\!\min\limits_{\substack{n\in \{1,\cdots,M/2\}\\s \in \mathcal{S}}}\!
\left(2\,(1-\Re\{s\,\e_1^H\M^{n}\e_1\})\right)^\frac{1}{N}\!\!\!\!.
\label{COH:propose:G2}
\end{equation}

Thanks to equation \eqref{COH:min:coding:gain}, problem \eqref{eq:maxge1} can be significantly simplified as follows.

\begin{table*}[t]
\centering
\caption{Examples of angle vectors of the optimal bases}
\begin{tabular}{|c|c|}\hline
\textbf{Projection Vector} & \textbf{Associated Optimal Angle Vector}\\
$\e_1(M,M)$ &  [$\theta_1~\theta_2~\cdots~\theta_M$] [rad] \\ \hline
$\e_1(2,2)$ & [0~0.785] \\ \hline
$\e_1(4,4)$ & [0~5.035~5.497~1.108] \\ \hline
$\e_1(8,8)$ & [0~0.930~0.022~0.243~4.180~1.367~0.376~2.481] \\ \hline
$\e_1(16,16)$ & [0~2.863~0.526~1.600~0.540~4.089~2.253~1.462~4.388~0.516~0.094~0.784~3.643~4.704~5.234~3.842] \\ \hline
$\e_1(32,32)$ &
\begin{tabular}{l}
[0~1.797~3.056~4.042~0.800~2.662~3.432~1.046~5.030~1.771~0.664~2.501~0.211~0.220~4.908~2.929\\~0.570~5.512~5.216~0.041~2.159~0.818~2.738~0.301~0.999~6.158~0.972~2.938~3.659~3.194~3.054~1.186]
\end{tabular} \\ \hline
$\e_1(64,64)$ &
\begin{tabular}{l}
[0~2.558~1.212~2.052~0.222~4.775~0.340~1.963~4.732~1.887~3.042~2.838~5.219~2.947~2.479~2.616\\~0.886~0.388~5.818~1.125~2.561~2.520~4.397~0.534~2.633~4.189~4.222~5.149~4.431~2.878~5.046~1.110\\~0.428~1.453~4.907~4.505~3.310~5.717~0.370~2.818~1.351~2.558~2.651~5.939~1.137~3.975~2.368~1.186\\~4.512~2.558~3.013~5.677~2.558~0.134~6.088~0.383~0.628~4.088~1.178~5.574~2.614~0.453~0.180~5.205]
\end{tabular} \\ \hline
\end{tabular}
\label{tab:bases:param}
\end{table*}

First, notice that $g_1(\e_1)$ is not actually a function of $\e_1$ and therefore is irrelevant to the maximization of coding gain via selection of the projection vector $\e_1$ as per equation \eqref{eq:maxge1}.
In turn, maximizing $g_2(\e_1)$ directly is challenging, because the term $\Re\{s\,\e_1^H\M^{n}\e_1\}$ rotates both with $n$ and $s$, such that a given $L$-PSK symbol $s$ that minimizes $\Re\{s\,\e_1^H\M^{n}\e_1\}$ for a given value of $n$, maximizes the same term for another value of $n$ corresponding to a rotation of $\pi/2$ radians, and vice-versa.

In light of the above, and noticing that the term $\{s\,\e_1^H\M^{n}\e_1\}$ is a complex number of radius $|\e_1^H\M^{n}\e_1|$, it follows that a robust strategy to maximize $g_2(\e_1)$ for all values of $n$ and $s$ is to minimize the average amplitude of the term $\e_1^H\M^{n}\e_1$ over $n$, such that the following alternative problem may be proposed as a relaxation of problem \eqref{eq:maxge1}
%
\begin{equation}
\begin{aligned}
\minimize_{\mathbf{\e_1}\in \mathbb{C}^{M \times 1}} \quad & f(\e_1) \\
\textrm{s.t.} \quad & |e_m|^2=1,\;\forall\; m=\{1,\cdots,M\},
\end{aligned}
\label{eq:minfe1}
\end{equation}
where the objective function $f(\e_1)$ is defined as
\begin{equation}
f(\e_1) \triangleq \sum_{n=1}^{M/2}|\e_1^H \M^n \e_1|.
\label{COH:propose:fanc}
\end{equation}

Unlike the original problem \eqref{eq:maxge1}, which is highly complex and dependent on a combinatorially large number of evaluation of a cost function of cubic complexity order $\mathcal{O}(M^3)$, the latter problem \eqref{eq:minfe1} is, although not convex, a continuous manifold optimization problem, based on an objective of complexity order $\mathcal{O}(M^2)$, which can be solved efficiently.
In fact, in some cases, solutions of problem \eqref{eq:minfe1} can be obtained in closed form, as is the case for $(M,L)=(2,4)$.
Specifically, in this case we can set $\e_1 = \left[e^{j \theta_1} ~ e^{j \theta_2}\right]^T / \sqrt{2}$, without loss of generality, and search for the pair $(\theta_1, \theta_2)$ that minimize $f(\e_1)$, which leads to the solution $(\theta_1, \theta_2) = (0, \pi/4)$.
This solution yields the optimum projection vector $\e_1=[1~e^{j\frac{\pi}{4}}]^T/\sqrt{2}\in \mathbb{C}^{2 \times 1}$, which in fact results in $f(\e_1) = 0$.

In the more general case, the optimization problem \eqref{eq:minfe1} can be solved via any number of publicly available standard optimization methods such as COBYLA, SLSQP \cite{kraft_software_1988}, BFGS, L-BFGS, and Powell \cite{buhmann_michael_2019}.
The performance of these methods will be compared in Subsection \ref{subsec:numerical}, and examples of angle vectors resulting in optimal $\e_1$ designs for various values of $M$ are given in Table \ref{tab:bases:param}.

\subsection{Extension to Sparse Projection Vectors and Matrices}
\label{subsec:extension}

Notice that the projection vector design of Subsection \ref{subsec:prop:basis}  generally yields fully dense solutions ($i.e., |\e_1|_0 = M$).
Allowing for sparsity in the projection may be desired, especially in large scale MIMO systems, as it enables a reduction of the number of RF chains required to implement the scheme.
In addition, it is also know that the coding gain of ADSM-based schemes increases with sparsity \cite{rajashekar_algebraic_2017} in the projection.
These facts motivate us to seek a sparse alternative to the design described in  Subsection \ref{subsec:prop:basis}.

On the other hand, recall that the utilization of a projection vector $\e_1$, as opposed to a projection matrix $\E_1$, represents already a significant compression of transmission instances to $T=1$.
With the additional insertion of sparsity into $\e_1$, the total number of degrees of freedom of the system is further reduced, which therefore motivates us also to extend the design of Subsection \ref{subsec:prop:basis} to the case when $T>1$.

These two modifications are the objectives of this subsection.
For the sake of clarity, we shall hereafter denote the original projection vector $\e_1$ of Subsection \ref{subsec:prop:basis} by $\e_1(M, M)$, so as to explicitly indicate that it is designed for $M$ transmit antennas, with $M$ nonzero entries.
Accordingly, a sparse projection vector with $|\e_1|_0 = N_b \leq M$ shall be denoted by $\e_{1} (M,N_b) \in \mathbb{C}^{M \times 1}$, which can be constructed as follows
\begin{align}
\e_1(M, N_b)=\e_1 \left(\frac{M}{2}, N_b \right) \otimes \begin{bmatrix}1 \\ 0\end{bmatrix} \in \mathbb{C}^{M \times 1}.
\label{COH:expand:basis}
\end{align}

For example, if we consider the case with $M=4$ and $N_b=2$, the corresponding extended basis is given by
\begin{equation}
\e_1(4, 2) \!\,= \!\,\e_1(2, 2) \otimes\! \begin{bmatrix}1 \\ 0\end{bmatrix}
\!\!=\!\! \frac{1}{\sqrt{2}}\! \begin{bmatrix}1 \\ e^{j\frac{\pi}{4}}\end{bmatrix} \!\otimes\! \begin{bmatrix}1 \\ 0\end{bmatrix}
\!\!=\!\!\frac{1}{\sqrt{2}}\! \begin{bmatrix} 1 \\ 0 \\ e^{j\frac{\pi}{4}} \\ 0 \end{bmatrix}\!\!.
\end{equation}

The sparse projection vector described by equation \eqref{COH:expand:basis}  is shown in Appendix \ref{app:gain_sparse_e1} to retain the same coding gain as that maximized under the solution of problem \eqref{eq:minfe1}, and therefore can be said to be the optimal sparse solutions of the latter.

Finally, a projection vector $\e_1(M, N_b)$ can be extended into a projection matrix denoted by $\E_1(M,N_b,T)\in \mathbb{C}^{M \times T}$, which can be constructed as
\begin{eqnarray}
\label{COH:expand:multi:T}
\E_1(M,N_b,T)=&&\\
&&\hspace{-16ex}\begin{pmat}[{|||}]\I_M\e_1(M,N_b) & \P\e_1(M,N_b) & \cdots &\P^{(T-1)}\e_1(M,N_b)\cr\end{pmat},
\nonumber
\end{eqnarray}
where $\P$ is a permutation matrix defined as
\begin{equation}
\P\triangleq\begin{pmat}[{|}]
\mathbf{0}_{1\times (M-1)} & 1 \cr
\-
\I_{M-1} & \mathbf{0}_{(M-1)\times 1} \cr
\end{pmat}.
\label{eq:pmatrix}
\end{equation}

Notice that left-multiplication of the matrix $\P^{n}$ onto a vector $\mathbf{v}$ yields the $n$-rotation of the elements of $\mathbf{v}$ in the downward direction.

For example, if $\mathbf{v}=[1~2~3~4]^T$, then $\P\mathbf{v}=[4~1~2~3]^T$, $\P^2\mathbf{v}=[3~4~1~2]^T$, and so on.
It follows that the projection matrix designed via equation \eqref{COH:expand:multi:T} consists of column-wise rotations of the design of equation \eqref{COH:expand:basis}.
For example, with $M=4,N_b=2$, and $T=2$ we obtain
\begin{align}
\E_1(4,2,2)=\begin{pmat}[{|}]\I_4\e_1(4,2)&\P\e_1(4,2)\cr
\end{pmat}=\frac{1}{\sqrt{2}}\left[\begin{array}{cc}
	1 & 0 \\
	0 & 1 \\
	e^{j\frac{\pi}{4}} & 0\\
	0 & e^{j\frac{\pi}{4}}
	\end{array}\right].
\end{align}

\subsection{Numerical Evaluation of Otimization Methods
\label{subsec:numerical}}

As mentioned above, the design of dense projection vectors based on the solution of problem \eqref{eq:minfe1} can be accomplished by standard optimization methods such as COBYLA, SLSQP, BFGS, L-BFGS, and Powell \cite{kraft_software_1988, buhmann_michael_2019}.
In this subsection we offer a numerical comparison of the performance of these methods.

\begin{figure}[tb]
\centering
\includegraphics*[width=0.92\columnwidth]{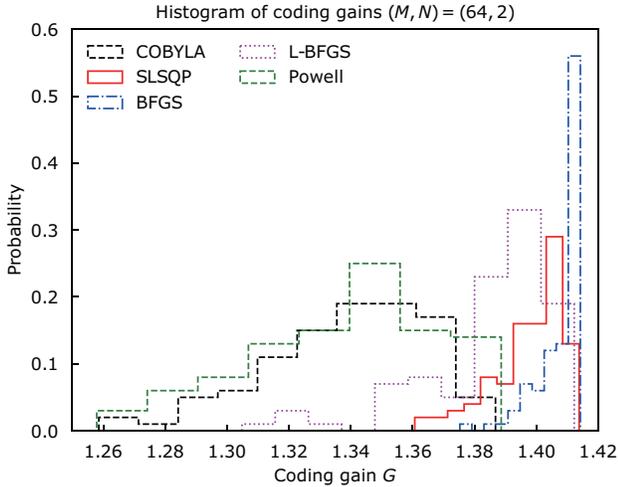}
\caption{Histogram of coding gains $g(\e_1)$ as given by equation \eqref{COH:coding:gain}, achieved by projection vectors obtained by solving problem \eqref{eq:minfe1} via various optimization packages, with $M=64$ and $N=2$.
\label{fig:optimization:M=64}}
\end{figure}
We start with Fig. \ref{fig:optimization:M=64}, which shows histograms of the coding gains $g(\e_1)$ obtained from equation \eqref{COH:coding:gain} with dense projection vectors $\e_1$ obtained by solving problem \eqref{eq:minfe1} via various optimization methods, for systems with $M=64$ transmit and $N=2$ receive antennas.
It is found that all methods yield gains $g(\e_1)> 1.26$, with the BFGS method proving most effective in achieving the highest gains, $g(\e_1) \approx 1.41$, with the highest probabilities, followed closely by the SLSQP method.

\begin{figure}[tb]
\includegraphics*[width=0.92\columnwidth]{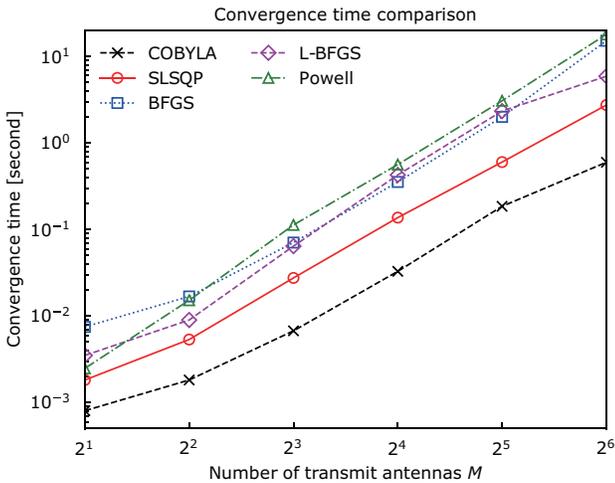}
\caption{Convergence time of solutions of problem \eqref{eq:minfe1} obtained via various optimization packages with the same initial value, as a function of system dimension $M = 2 \to 64$.
\label{fig:optimization:time}}
\end{figure}
Next, we compare the convergence time of each of the latter optimization methods as a function of the system size.
The results are shown in Fig. \ref{fig:optimization:time} and indicate that all methods are relatively fast, with the solutions for systems with $M = 64$ antennas obtained in around 10 seconds with the slowest approach (Powell method), and in less than 1 second with the fastest (COBYLA method).
Considering both sets of results, we hereafter adopt the SLSQP method as it is found to be both fast and effective in achieving a high coding gain with a high probability.

\begin{figure}[tb]
\includegraphics*[width=0.92\columnwidth]{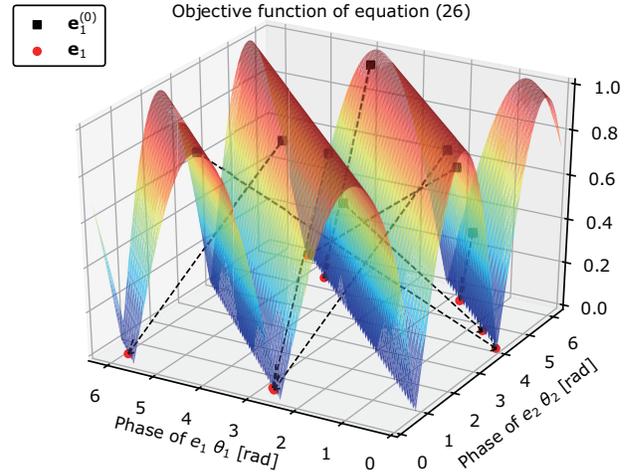}
\caption{Starting and converging points of problem \eqref{eq:minfe1} with $M=2$.
\label{fig:3D}}
\end{figure}
Finally, Fig. \ref{fig:3D} shows the initial and converged points of the optimization problem \eqref{eq:minfe1} solved by SLSQP method, with $M=2$.
It can be seen that the objective function has many local minima, such that the final convergence point is highly distinct. Some initial points converge to the nearest minima, some do not, which makes it difficult for eavesdroppers to estimate the initial value from the converged value, even if $M=2$.

\section{N-DSTC with Embedded One-way like function Physical Layer Security}
\label{sec:ppro}

The proposed N-DSTC scheme described above also has the feature that the projection matrix $\E_1$ is obtained from the expansion via equations \eqref{COH:expand:basis} and \eqref{COH:expand:multi:T}, of the solution problem \eqref{eq:minfe1} which, due to its manifold nature and the non-convexity of its objective, is not unique but rather dependent both on the method employed and on the initial value of the iterations towards a stable point, as illustrated in Figs. \ref{fig:optimization:M=64} and \ref{fig:3D}.

This feature is similar to the notion of one-way function\footnote{A one-way function is defined as a function whose output is easy to compute for all inputs, but whose inverse is hard. However, no one has yet been able to answer whether there exists a one-way function in the sense of cryptographic theory.} \cite{chhabra_design_2020} widely used in cryptographic algorithms, and therefore can be exploited to embed physical layer security into the proposed N-DSTC scheme\footnote{This requires that the space-time modulation scheme generates symmetric symbols, such as in the ADSM approach introduced in Section~\ref{subsec:stm}.}, as described below.

\subsection{Proposed PLS-protected N-DSTC Scheme}
\label{sec:pproalgo}

Straightforwardly, the proposed N-DSTC scheme can be summarized as follows:

\begin{enumerate}
\item Assuming that the transmitter Alice and the receiver Bob share a secret key\footnote{For a full physical-layer implementation, $z$ can be assumed to be extracted from the channel between Alice and Bob \cite{allen_secure_2014}.}, $z \in \mathbb{Z}$, the pair locally generates an identical pseudo-random initial vector $\e_1^{(0)}\in \mathbb{C}^{M \times 1}$ based on the seed $z$.

\item Starting from $\e_1^{(0)}$, Alice and Bob locally solve problem \eqref{eq:minfe1} using the same optimization method and parameters ($e.g.$ number of iterations, step sizes, etc.) which may also be pseudo-randomly determined by the shared seed $z$, thus obtaining solution vector $\e_1$ common to the pair.
\item Using the common vector $\e_1$, Alice and Bob locally produce $\E_1$ via equations \eqref{COH:expand:basis} and \eqref{COH:expand:multi:T}.

\item During the first $T$ instances Alice transmits a random unitary matrix $\U_M=[\U_1, \cdots, \U_{M/T}]\in \mathbb{C}^{M \times M}$, subsequently transmitting its payload data matrices $\X(i)$, differently encoded and encrypted by $\E_1$, yielding at Bob the received symbol blocks
\begin{equation}
\Y(i)\!=\!
\begin{cases}
\H(i)\U_i + \V(i) & \!\!\text{for } 1 \!\leq\! i \!\leq\! M/T,\\
\H(i)\underbrace{\X(i\!-\!1)\X(i)}_{\triangleq \S(i)}\E_1\!+\!\!\V(i) & \!\!\text{for } i > M/T,\\[-3ex]
\end{cases}
\label{COH:proposed_transmitter}
\end{equation}
where $\S(i)$ are the differentially-encoded symbol blocks equivalent to those constructed via equation \eqref{COH:differencial} in conventional schemes, and $\X(M/T)\triangleq \I_M$.

\item Bob differentially decodes and decrypts the messages from Alice
by computing
\begin{equation}
\Xh(i)=\argmin_{\X} ||\Y(i)-\Yh(i-1)\,\X\,\E_1||^2_F,
\label{COH:propose:nonsq:detection}
\end{equation}
with $\Yh(i)$ defined as
\begin{equation}
\Yh(i) \!\triangleq\!
\begin{cases}
\sum\limits_{k=1}^{M/T}\! \Y(k)\U^H_k & \!\!\!\text{for } i\!=\!M/T,  \\
\Y(i)\E^{(1-\alpha)}\!+\!\!\Yh(i\!-\!1)\Xh(i)\E^{(\alpha)} & \!\!\!\text{for } i\!>\!M/T,
\end{cases}
\label{COH:propose:Y:hat}
\end{equation}
the constant matrices $\E^{(\alpha)}$ and $\E^{(1-\alpha)}$ defined as
\begin{subequations}
\label{COH:propose:constant:matrix}
\begin{eqnarray}
&\E^{(\alpha)}\triangleq\I_M-\E_1\E^{(1-\alpha)},&\\
&\E^{(1-\alpha)}\triangleq(1-\alpha(i))\E_1^H,&
\end{eqnarray}
\end{subequations}
and $\alpha(i)$ as given in equation \eqref{COH:forgetting:factor}.

\item Restart the procedure from step 1) periodically.

\end{enumerate}

Notice that unlike conventional schemes reviewed in Subsection \ref{subsec:basis}, the proposed scheme does not rely on an entire set of compression matrices $\E_1,\E_2, \cdots, \E_{M/T}$, but rather on a single projection/encryption matrix $\E_1$.
This feature not only simplifies the ML detection process, as can be inferred by direct comparison of equations \eqref{COH:constant:matrix} and \eqref{COH:propose:constant:matrix}, but also is crucial for security, since $\E_1$ is uniquely determined by $\E_2, \cdots, \E_{M/T}$.

Furthermore, also unlike conventional schemes, the secret projection/encrypting matrix $\E_1$ is never exposed to Eve, nor is it related to the unitary matrix transmitted during the initialization phase of the differentially encoding procedure, in contrast to conventional schemes, $e.g.$ as in equation \eqref{eq:unitary_to_E}.

In light of the features highlighted above, in order to decrypt the secret messages exchanged between Alice and Bob, an eventual eavesdropper Eve must extract $\E_1$ directly from its own received signals, subject to a distinct channel.
In the next subsections, we formulate the best attack Eve can inflict onto the proposed scheme and subsequently analyze the secrecy rate of the method.

\subsection{Eavesdropping Attack Conditions and Strategy}
\label{sec:attackalgo}

In order to assess the secrecy performance of the N-DSTC scheme with embedded one-way  function physical layer security under the most strenuous conditions, an idealized scenario highly beneficial to Eve is considered, namely:
\begin{itemize}
\item {\bf Perfect CSI}: the channel matrix $\H_{\mathrm{Eve}}$ from Alice to Eve is assumed to be perfectly known, despite the fact that the N-DSTC scheme does not incorporate any procedure for channel estimation.
\item {\bf Noise-free Prior Information}: it is assumed that Eve is in knowledge of the first transmitted message $\tilde{\X}\triangleq\X(M/T+1)$ and in possession of a noise-free copy of the corresponding received signal, namely
\begin{equation}
\label{eq:EvePriorRXSignal}
\tilde{\Y}_{\mathrm{Eve}} = \H_{\mathrm{Eve}} \tilde{\X} \E_1.
\end{equation}
\item{\bf Coherent Detection:} combining the perfect CSI assumption with the prior information summarized in equation \eqref{eq:EvePriorRXSignal}, and in view of the N-DSTC transmission model of equation \eqref{COH:proposed_transmitter}, it is assumed that Eve can attempt to detect Alice's messages coherently from the signal model
\begin{equation}
\label{eq:EveRXSignal}
\Y_{\mathrm{Eve}}(i)=\H_{\mathrm{Eve}}\,\X(i) \, \E_1 + \V_{\mathrm{Eve}}(i).
\end{equation}

\item {\bf System Information}: it is assumed that Eve is in full knowledge of the fact that Alice and Bob generate $\E_1$ via the solution of problem \eqref{eq:minfe1} with the objective function given in equation \eqref{COH:propose:fanc}.
\end{itemize}

Under the highly favorable conditions described above, Eve can mount the following idealized attack\footnote{Referring to equation \eqref{COH:propose:basis}, notice that the search space of $\E_1$ is of dimension $2^{\beta M}$, where $\beta$ denotes the resolution of $\theta_m$, and that under practical conditions Eve must also estimate $\H_{\mathrm{Eve}}$ and infer $\X(M/T+1)$ while searching for $\E_1$, making brute-force approaches virtually impossible with sufficiently large $\beta$ and/or $M$.} aiming at extracting the encrypting projection matrix $\E_1$
\begin{equation}
\hat{\E}_1\!=\!\argmin_{\E_{1_{\mathrm{Eve}}}}\!\big\|\tilde{\Y}_{\mathrm{Eve}}-\H_{\mathrm{Eve}}\tilde{\X}\E_{1_{\mathrm{Eve}}}\big\|^2_F+\!\sum_{n=1}^{M/2}\!\big|\E_{1_{\mathrm{Eve}}}^H \M^n \E_{1_{\mathrm{Eve}}}\big|,
\label{COH:Attacking2}
\end{equation}
and subsequently attempt to decipher $\X(i)$ by solving
\begin{equation}
\Xh(i)=\argmin_{\X} \big\|\Y_{\mathrm{Eve}}(i)-\H_{\mathrm{Eve}}\X\hat{\E}_1\big\|^2_F, \;\text{for } i > M/T.
\label{COH:cohe:nonsq:detection}
\end{equation}

\subsection{Secrecy Rate Analysis}

The resilience of the proposed PLS-protected N-DSTC scheme described in Subsection \ref{sec:pproalgo} to the eavesdropping attack described above can be assessed via its secrecy rate under a finite-alphabet signaling which is given by \cite{wang_secrecy_2015,rezaei_aghdam_overview_2019,shu_two_2018,jiang_secrecy-enhancing_2018}
\begin{equation}
C=\mathrm{max}\{0,(I_{\mathrm{Bob}}-I_{\mathrm{Eve}})\},
\label{COH:secret}
\end{equation}
where $I_{\mathrm{Bob}}$ and $I_{\mathrm{Eve}}$ denote the average mutual information (AMI) between the information source Alice, and Bob or Eve, respectively, which are given by \cite{soon_xin_ng_mimo_2006}
\begin{equation}
I_{\mathrm{Bob}}\!=\!\frac{1}{T}\!\left(\!B\!-\!\frac{1}{L}\!\sum_{p=1}^{L} \mathrm{E}_{\H,\!\V}\mathrm{log}_2\!\left(\sum_{q=1}^{L}\frac{\mathrm{p}(\Y_{\mathrm{Bob}}^{(p)}|\X^{(q)}\E_1)}{\mathrm{p}(\Y_{\mathrm{Bob}}^{(p)}|\X^{(p)}\E_1)}\right)\!\!\right)\!,
\label{COH:def:AMIBob}
\end{equation}
with $\Y_{\mathrm{Bob}}$ simplified, for analytical purposes to
\begin{equation}
\label{eq:BobRXSignalCoherent}
\Y_{\mathrm{Bob}}(i)=\H\X(i)\E_1+\V(i),
\end{equation}
and
\begin{equation}
I_{\mathrm{Eve}}\!=\!\frac{1}{T}\!\left(\!\!B\!-\!\frac{1}{L}\!\sum_{p=1}^{L} \mathrm{E}_{\H_{\mathrm{Eve}},\!\V_{\mathrm{Eve}}}\mathrm{log}_2\!\left(\sum_{q=1}^{L}\!\frac{\mathrm{p}(\Y_{\mathrm{Eve}}^{(p)}|\X^{(q)}\hat{\E}_1)}{\mathrm{p}(\Y_{\mathrm{Eve}}^{(p)}|\X^{(p)}\hat{\E}_1}\!\right)\!\!\right)\!,
\label{COH:def:AMIEve}
\end{equation}
where we remind that $L \triangleq 2^B$, and the auxiliary indices $p$ and $q$, with $1 \leq p,q\leq L$ are used to indicate distinct symbol matrices in the codebook $\mathcal{X}\triangleq \{\X_1, \cdots, \X_{2^B}\} \in \mathbb{C}^{M \times M}$.

In light of equations \eqref{eq:EveRXSignal} and \eqref{eq:BobRXSignalCoherent}, the conditional probabilities in equations \eqref{COH:def:AMIBob} and \eqref{COH:def:AMIEve} are respectively given by
\begin{equation}
\mathrm{p}(\Y_{\mathrm{Bob}}|\X\E_1)\!=\!\frac{1}{(\pi\sigma_v^2)^{NT}}\mathrm{exp}
\!\left(\!\frac{-\|\Y_{\mathrm{Bob}}\!\!-\!\H\X\E_1\|^2_F}{\sigma_v^2}\!\right)\!,
\label{COH:def:pBob}
\end{equation}
%
\begin{equation}
\mathrm{p}(\Y_{\mathrm{Eve}}|\X\hat{\E}_1)\!=\!\frac{1}{(\pi\sigma_{v:\text{Eve}}^2)^{NT}}\mathrm{exp}
\bigg(\!\frac{-\|\Y_{\mathrm{Eve}}\!-\!\H_{\mathrm{Eve}}\X\hat{\E}_1\|^2_F}{\sigma_{v:\text{Eve}}^2}\!\bigg).
\label{COH:def:pEve}
\end{equation}

Substituting equations \eqref{COH:def:pBob} and \eqref{COH:def:pEve} into \eqref{COH:def:AMIBob} and \eqref{COH:def:AMIEve}, respectively, yields
\begin{equation}
I_\text{Bob}=\frac{1}{T} \bigg(
B-\frac{1}{L}\sum_{f=1}^{L} \mathrm{E_{\H,\V}}\Big[
\mathrm{log}_2\sum_{g=1}^{L}e^{\eta^{(p,q)}_\text{Bob}}
\Big]\bigg),
\label{COH:def:AMI:2Bob}
\end{equation}
\begin{equation}
I_\text{Eve}=\frac{1}{T} \bigg(
B-\frac{1}{L}\sum_{f=1}^{L} \mathrm{E_{\H_\text{Eve},\V_\text{Eve}}}\Big[
\mathrm{log}_2\sum_{g=1}^{L}e^{\eta^{(p,q)}_\text{Eve}}
\Big]\bigg),
\label{COH:def:AMI:2Eve}
\end{equation}
where
\begin{equation}
\eta_{\mathrm{Bob}}^{(p,q)}\triangleq\frac{-||\H(\X^{(p)}-\X^{(q)})\E_1+\V||^2_F}{\sigma_v^2},
\label{COH:def:eta:Bob}
\end{equation}
\begin{align}
\eta_{\mathrm{Eve}}^{(p,q)}\triangleq &\; \frac{-\|\H_{\mathrm{Eve}}(\X^{(p)}\E_1-\X^{(q)}\hat{\E}_1)+\V_{\mathrm{Eve}}\|^2_F}{\sigma_{v:\text{Eve}}^2}\nonumber\\&+\frac{\|\H_{\mathrm{Eve}}\X^{(p)}(\E_1-\hat{\E}_1)+\V_{\mathrm{Eve}}\|^2_F}{\sigma_{v:\text{Eve}}^2}.
\label{COH:def:eta:Eve}
\end{align}

\section{Simulation Results}
\label{sec:comp}

In this section, we evaluate the performance of the proposed scheme, first in terms of the communication between Alice and Bob, and then in terms of its security, in particular in terms of the leakage in the Alice-to-Eve channel.

\subsection{Performance Comparison}
\label{subsec:performance}

In this subsection we compare the proposed nonsquare coding (ADSM with the proposed projection vector/matrix designed in Subsections \ref{subsec:prop:basis} and \ref{subsec:extension}) and the conventional nonsquare coding (ADSM and DUC with the conventional projection vector/matrix designed in Subsection \ref{subsec:basis}) in terms of the coding gain and the bit error rate (BER).

\begin{figure}[tb]
\centering
\includegraphics*[width=0.92\columnwidth]{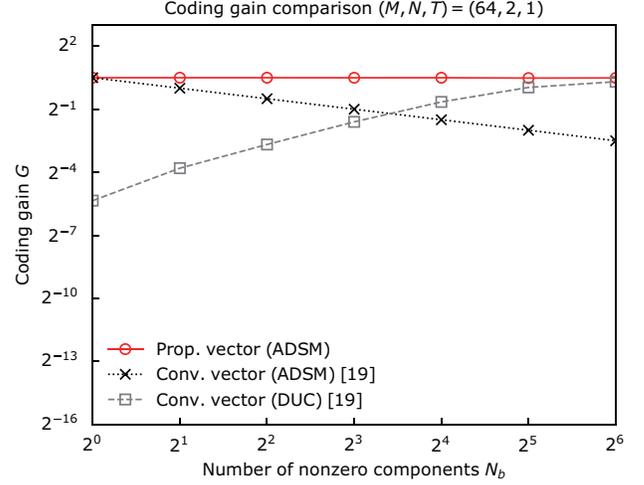}
\caption{Comparison of coding gains achieved as per equation \eqref{COH:coding:gain}, with projection vectors designed as given by result of optimization problem \eqref{eq:minfe1}, $M=64$, $N=2$, $T=1$, $B=8$ and $N_b = 1 \to 64$.}
\label{fig:compare:coding:gain}
\end{figure}
Fig. \ref{fig:compare:coding:gain} compares the coding gain achieved with projection vectors obtained from equation \eqref{COH:expand:basis}, systems with parameters $M=64, N=2, T=1$, and $B=8$, as a function of the sparsity of the vector, ranging from $N_b=1$ to $64$.
As references, curves corresponding to the conventional nonsquare ADSM and DUC scheme are also included\footnote{In the DUC case, the factors $u_1, u_2, \cdots ,u_M$ are optimized as described in \cite{hochwald_differential_2000} and indicated by equation \eqref{COH:DUC:fanc}, but only $N_b$ elements are utilized.}.

It can be seen that the coding gain of conventional nonsquare ADSM schemes decreases as the projection vector becomes more dense, with the opposite occurring in DUC approach.
In turn, the coding gain achieved using projection vectors designed via the method proposed in Subsections \ref{subsec:prop:basis} and \ref{subsec:extension} is found to be independent of sparsity, and to upper-bound those achieved by the latter schemes.

\begin{figure}[tb]
\includegraphics*[width=0.92\columnwidth]{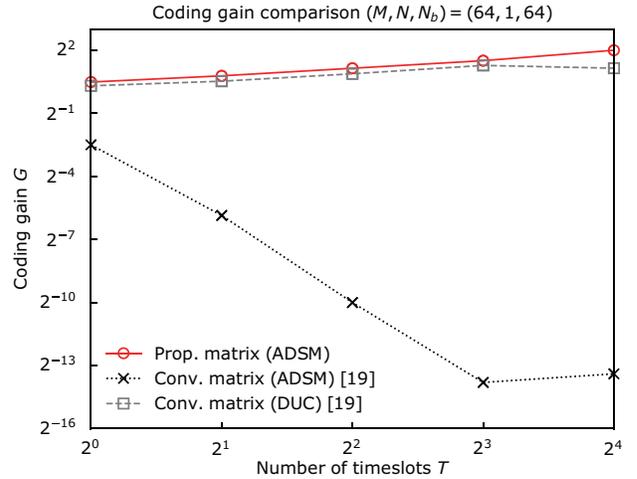}
\caption{Comparison of coding gains achieved as per equation \eqref{COH:coding:gain}, with projection matrices designed as given by equation \eqref{COH:expand:multi:T}, $M=64$, $N=2$, $B=8$, $N_b=64$ and $T = 1 \to 16$.}
\label{fig:compare:coding:gain:multi:T}
\end{figure}
Fig. \ref{fig:compare:coding:gain:multi:T} compares the coding gain achieved with projection matrices obtained from equation \eqref{COH:expand:multi:T}, systems with parameters $M=64, N=2, N_b=64$, and $B=8$, as the number of timeslots, ranging from $T=1$ to $16$.
Notice that with such parameterization, the matrices are actually dense.
It can be seen that in this case the proposed dense matrix design via \eqref{COH:expand:multi:T} achieves max coding gain, still outperforming the DUC approach.
It is also seen, furthermore, that the gain achieved under maximum compression ($i.e.$, with $T=1$) is not much lower than that achieved with $T=16$, which, taken together with the implications of the value of $T$ onto the rate of the system, indicates the desirability of designs with the $T\to 1$.
The latter feature will also be exploited in Section \ref{sec:ppro}, where the proposed scheme is further extended to incorporate physical layer security.

\begin{figure}[tb]
\centering
\includegraphics*[width=\columnwidth]{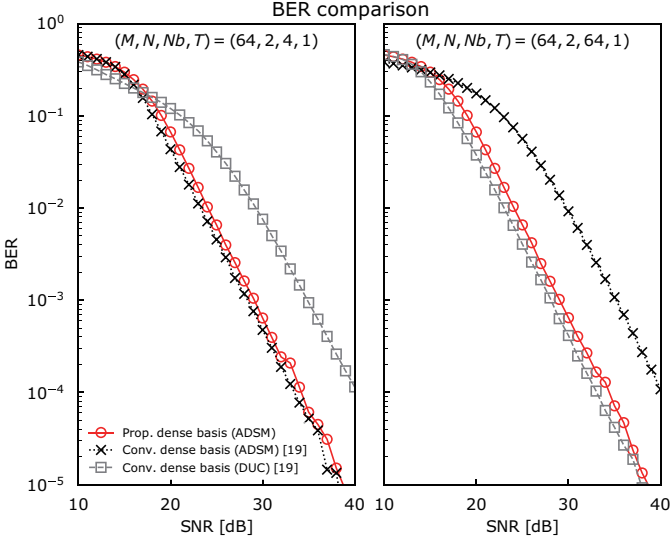}
\caption{Comparison of BERs achieved with projection vectors designed via optimization problem \eqref{eq:minfe1} and equation \eqref{COH:expand:basis}, with $M=64$, $N=2$, $N_b=4$ or $64$, $T=1$ and $B=8$.
\label{fig:compare:M64:BER}}
\end{figure}
Next, Fig.~\ref{fig:compare:M64:BER} shows the BER comparison of the proposed and conventional codings, system with parameters $M=64,N=2,N_b=4$ or $64,T=1$, and $B=8$.
The results corroborate those of Figs. \ref{fig:compare:coding:gain} and \ref{fig:compare:coding:gain:multi:T}, by showing that the BER performance of the proposed schemes is, unlike that of conventional methods, independent of sparsity.

\begin{figure}[tb]
\includegraphics*[width=0.92\columnwidth]{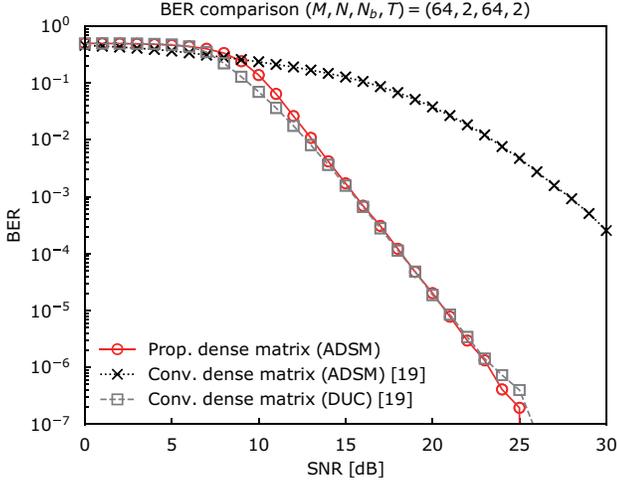}
\caption{Comparison of BERs achieved with projection matrices designed via equation \eqref{COH:expand:multi:T}, with $M=64$, $N=2$, $N_b=64$, $T = 2$ and $B=8$.
\label{fig:compare:BER:T=2}}
\end{figure}
Finally, Fig. \ref{fig:compare:BER:T=2} shows the BER comparison of the proposed and conventional schemes, system with parameters $M=64,N=2,N_b=64,T=2$, and $B=8$. It is seen that the proposed method achieves the same performance as the DUC, despite the significantly lower complexity. Notice that the BER of the proposed method, with $T=2$ is much better than that with $T=1$, however, the transmit rate $R \triangleq B/T~\text{[bit/symbol]}$ decreases from $R=8$ to $R=4$. These results reflect the results of Fig. \ref{fig:compare:coding:gain:multi:T} for the $T=2$ case. Further BER performance improvement can be obtained, albeit at the sacrifice of transmission rate, by setting $T > 2$.

\subsection{Security Performance Evaluation}

For the benefit of the eavesdropper, we evaluate in this subsection the secrecy performance of the proposed N-DSTC scheme with embedded security with parameters $N_b=M$ and $T=1$, since the sparsity resulting from setting $N_b < M$ and $T > 1$ on the one hand was shown in Subsection \ref{subsec:performance} not to negatively impact the BER performance of the proposed scheme, and on the other hand only increases the complexity of solving equation \eqref{COH:Attacking2}.
In addition, in what follows, we denote the number of receive antennas at Bod and Eve respectively by $N_{\mathrm{Bob}}$ and $N_{\mathrm{Eve}}$.

\begin{figure}[tb]
\centering
\includegraphics*[width=0.92\columnwidth]{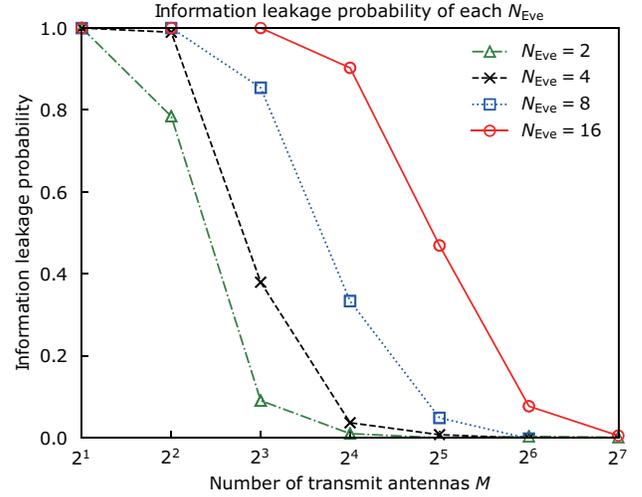}
\caption{Information leakage probability in an environment ideal to Eve,  with $N_{\mathrm{Eve}}=2,4,8,16$, $M=2 \to 128$ and $B=\mathrm{log}_2(M)+2$.}
\label{fig:Eve's:channel:reliability}
\end{figure}
Our first assessment, depicted in Fig. \ref{fig:Eve's:channel:reliability}, is the information leakage probability to Eve, measured as $1-2\times \mathrm{BER}_\text{Eve}$ \cite{yilmaz_relationships_2020}, with $\mathrm{BER}_\text{Eve}$ given by the average bits in error in the symbols $\Xh(i)$ detected by Eve under equation \eqref{COH:cohe:nonsq:detection}, using projection matrices $\hat{\E}_1$ estimated by solving equation \eqref{COH:Attacking2}, and given received signals $\tilde{\Y}_{\mathrm{Eve}}$ obtained in the absence of noise, as described by equation \eqref{eq:EvePriorRXSignal}. As can be seen from the definition of channel reliability [35], which in the context hereby can be understood as a leakage probability, when $\mathrm{BER}_\text{Eve}=0.5$ Eve's entropy becomes zero.

The figure shows plots of such leakage as a function of the number of transmit antennas $M$, varied from $M=2$ to $128$, and for different configurations receive antennas at Eve, varying from $N_{\mathrm{Eve}}=2$ to $N_{\mathrm{Eve}}=16$.

It can be seen that even under the highly idealized conditions assumed in favor of Eve, zero leakage can be achieved by the proposed scheme, as long as the number of the transmit antennas $M$ of the system is significantly larger than the number of receive antennas employed by Eve since $N_{\mathrm{Eve}}$ degrees of freedom are not sufficient for estimating the $M$ variables in $\e_1$.
We remark that the curves must be interpreted as fundamental bounds, not only because the assumption $\text{SNR}=\infty$ at Eve is impractical, but also because in realistic conditions it is very hard for Eve to estimate $\H_{\mathrm{Eve}}$ at all (let alone perfectly), since the N-DSTC scheme does not include a channel estimation procedure between Alice and Bob ($i.e.$, no pilot symbols are transmitted by the system).

\begin{figure}[tb]
\centering
\includegraphics*[width=0.92\columnwidth]{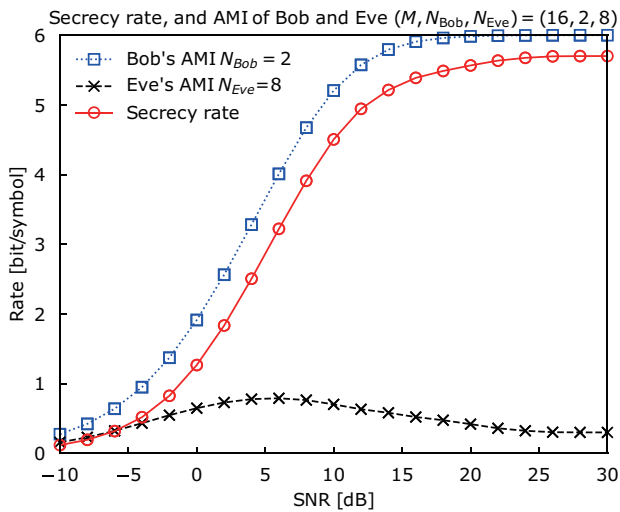}
\caption{Average mutual information from Alice to Bob and Eve, and corresponding secrecy rate as a function of SNR, with $M=16$, $N_{\mathrm{Bob}}=2$, $N_{\mathrm{Eve}}=8$ and $B=6$.
\label{fig:AMI}}
\end{figure}
Next, we compare in Fig. \ref{fig:AMI} average mutual information from Alice to Bob and Eve, respectively, as given by equations \eqref{COH:def:AMI:2Bob} and \eqref{COH:def:AMI:2Eve}, with $M=16,~N_{\mathrm{Bob}}=2,~N_{\mathrm{Eve}}=8$ and as a function of the SNR, assumed to be the same both at Bob and Eve.
For the sake of convenience, the secrecy rate obtained by equation \eqref{COH:secret} is also included.
The results highlight the consistency of the proposed secure N-DSTC scheme, as it can be seen that under the evaluated conditions Bob enjoys monotonically increasing AMI and secrecy rate, as SNR increases, with the maximum rate of 6 bit/symbol (determined by $B$) achieved at around $\text{SNR}=20$ dB.

\begin{figure}[tb]
\includegraphics*[width=0.92\columnwidth]{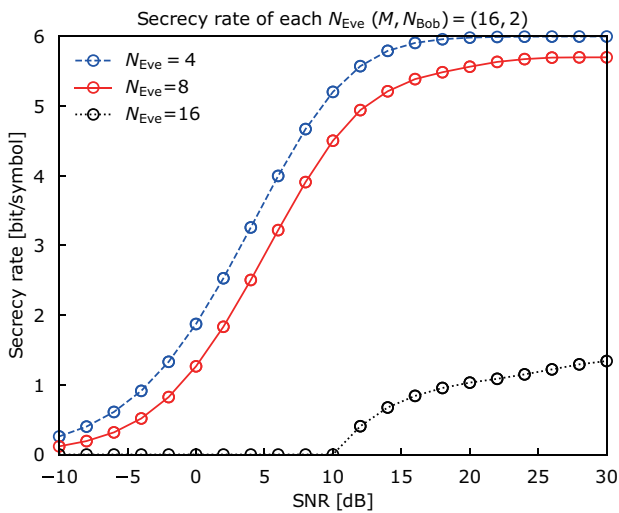}
\caption{Secrecy rate of a system with fixed transmit antennas $M=16$, legitimate receive antennas $N_{\mathrm{Bob}}=2$, in presence of eavesdroppers with different numbers of receive antennas $N_{\mathrm{Eve}}=4, 8, 16$, with $B=6$.}
\label{fig:Secret:M:16}
\end{figure}
Finally, we study in Fig. \ref{fig:Secret:M:16} the evolution of the secrecy rate achieved when the number of receive antennas employed by Eve increased to $N_{\mathrm{Eve}}=4,8,16$, in a system with a fixed number of transmit antennas, set to $M=16$, and of receive antennas employed by Bob, set to $N_{\mathrm{Bob}}=2$.
The results highlight the robustness of the scheme, as it can be seen that a secrecy rate equal to the achievable rate of the system, namely $B=6$ bit/symbol, can be reached at reasonable SNRs even when Eve has double the number of receive antennas as Bob, $i.e.$, $N_{\text{Eve}}=4$ and $N_{\text{Bob}}=2$, respectively.

In fact, it is found that even if Eve has four times the number of antennas employed by Bob ($N_{\text{Eve}}=8$ against $N_{\text{Bob}}=2$), the system still reaches a secrecy rate only a fraction of a bit away from the maximum achievable rate.
It is only when Eve employs as many receive antennas as the transmitter itself, $i.e.$ $N_{\text{Eve}}=M=16$, that the secrecy rate of the system drops to a fraction of the total achievable rate.

We remark that the fact that secrecy performance of the proposed secure N-DSTC is higher with increasing $M$, combined with the fact that the BER performance of the system is unaffected by sparsity in the projection matrices $\E_1$, make the contribution here presented particularly attractive to massive MIMO systems, since systems to a large number of transmit antennas and smaller number of RF chains can take full advantage of the embedded physical security provided by the proposed scheme, without any compromise in rate-performance.

\section{Conclusions}\label{sec:conc}
We proposed a novel nonsquare differential space-time coding MIMO scheme with embedded physical layer security.
In particular, in the proposed method, the space-time projection matrices characteristic of N-DSTC are constructed via the sparsification and column-wise expansion of vector solutions of a non-convex, continuous, coding gain maximization.
As a consequence of the non-convexity of the design problem and the use of a secret key private to Alice and Bob as initial value, the proposed projection matrices are similar to one way-functions, being difficult to be extracted by an eavesdropper even under highly favorable idealized conditions.
The proposed N-DTSC scheme is shown to improve upon the state of the art by providing embedded physical layer security, without any sacrifice in BER performance and with a significant reduction in the complexity of designing the projection matrices $\E_1$.
Since both the coding gain and the BER performance of the proposed scheme are shown to be independent of eventual sparsity of the projection matrices $\E_1$, and since the physical layer security achieved by the method improves with the number of transmit antennas irrespective of the sparse utilization of such resources.

\appendices
\section{Coding Gain in ADSM Case}
\label{app:coding:gain}

Referring back to the construction of data matrices in the ADSM scheme, as per equation \eqref{COH:ADSM:datamatrix}, let us express two generic and distinct ADSM data matrices by $\X_p=s_1\A_p$ and $\X_q=s_2\A_q$, with $1 \leq p,q \leq M$ and $s_1,s_2 \in \mathcal{S}$.
Substituting $\X_p=s_1\A_p$ and $\X_q=s_2\A_q$ into equation \eqref{COH:coding:gain}, the ADSM coding gain $g(\e_1)$ can be rewritten as follows
\begin{align}
g(\e_1)&=\!\! \min_{\X_p \neq \X_q\in \mathcal{X}}\!\! \left(\e_1^H(\X_p-\X_q)^H(\X_p-\X_q)\e_1 \right)^\frac{1}{N}\label{COH:A:eq:coding:gain}\\[-1ex]
&=\!\!\min_{{s_1, s_2 \in \mathcal{S}}\atop{\A_p,\A_q\in \mathcal{A}}} \!\!
\left(\e_1^H(s_1\A_p-s_2\A_q)^H(s_1\A_p-s_2\A_q)\e_1\right)^\frac{1}{N}\!\!\!.\nonumber
\end{align}

Notice that in order to have $\X_p \neq \X_q$, it is sufficient that either $s_1\neq s_2$, or $\A_p\neq\A_q$, or both.

Starting with the first case, $i.e.$ for $\A_p=\A_q=\A\in \mathcal{A}$ and thus $p=q$, and using the fact that $\mathcal{A}$ is a set of unitary matrices as evidenced by equations \eqref{COH:eq:ADSM:batrix} and \eqref{COH:eq:permutation:matrix}, equation \eqref{COH:A:eq:coding:gain} reduces to
\begin{align}
g(\e_1)&=\min_{{s_1, s_2 \in \mathcal{S}}\atop{\A\in \mathcal{A}}}
\left( \e_1^H(s_1-s_2)^*\A^H\A(s_1-s_2)\e_1\right)^\frac{1}{N}\nonumber\\[-1ex]
&=\min_{s_1, s_2 \in \mathcal{S}}
\left[ (s_1-s_2)(s_1-s_2)^*\e_1^H\e_1\right]^\frac{1}{N}\nonumber\\[-1ex]
&=\min_{s_1, s_2 \in \mathcal{S}}
\left|s_1-s_2\right|^\frac{2}{N}, \;\text{for } p=q,
\label{COH:propose:G1Appendix}
\end{align}
which is identical to equation \eqref{COH:propose:G1}.

Next, let us consider the case when $p \neq q$. For convenience, and without loss of generality, let us set $p < q$, such that equation \eqref{COH:A:eq:coding:gain} becomes
\begin{align}
g(\e_1)&=\!\!\!\!\!\!\min_{\substack{p\neq q\in \{1,\cdots,M\}\\s_1, s_2 \in \mathcal{S}}}\!
\left( \e_1^H(s_1\A_p\!-\!s_2\A_q)^H(s_1\A_p\!-\!s_2\A_q)\e_1\right)^\frac{1}{N}\nonumber\\
&=\!\!\!\!\!\!\min_{\substack{p\neq q\in \{1,\cdots,M\}\\s_1, s_2 \in \mathcal{S}}}\!
\left( \e_1^H(s_1^*\A_p^H-s_2^*\A_q^H)(s_1\A_p-s_2\A_q)\e_1\right)^\frac{1}{N}\nonumber
\end{align}
\begin{align}
\quad&=\!\!\!\!\!\!\min_{\substack{p\neq q\in \{1,\cdots,M\}\\s_1, s_2 \in \mathcal{S}}}\!
\left(\e_1^H(|s_1|^2\A_p^H\A_p-s_1^*s_2\A_p^H\A_q\right.\nonumber\\[-3ex]
&\hspace{15ex} \left.-s_1s_2^*\A_q^H\A_p+|s_2|^2\A_q^H\A_q)\,\e_1\right)^\frac{1}{N}\nonumber\\
&=\!\!\!\!\!\!\min_{\substack{p\neq q\in \{1,\cdots,M\}\\s_1, s_2 \in \mathcal{S}}}\!\!
\left(2\!-\!\e_1^H(s_1^*s_2\A_p^H\A_q\!\!+\!(s_1^*s_2\A_p^H\A_q)^H)\e_1\right)^{\!\frac{1}{N}}\!\!\!\!,
\label{COH:A:eq:coding:gain2}
\end{align}
where we have used the fact that $\mathcal{S}$ is an $L$-PSK constellation, such that $|s|^2=1,\;\forall\;s\in\mathcal{S}$.

Using also that fact that $\A_p^H\A_q=\M^{q-p}=\M^{n}$, where $n\triangleq q-p$ with $1 \leq n \leq M-1$, and given that $s_1^*s_2\in \mathcal{S}$, let us for notational convenience define $s \triangleq s_1^*s_2$ such that we may re-write equation \eqref{COH:A:eq:coding:gain2} as
\begin{align}
g(\e_1)
&=\!\!\!\!\!\!\min_{\substack{n\in \{1,\cdots,M-1\}\\s \in \mathcal{S}}}\!
\left( 2-\e_1^H(s\,\M^{n}+s^*(\M^{n})^H)\e_1\right)^\frac{1}{N}\label{COH:A:eq:coding:gain:AiAj}\\
&=\!\!\!\!\!\!\min_{\substack{n\in \{1,\cdots,M-1\}\\s \in \mathcal{S}}}\!
\left( 2-s\,\e_1^H\M^{n}\e_1+s^*\e_1^H(\M^{n})^H\e_1\right)^\frac{1}{N}\!\!\!\!.\nonumber
\end{align}

Next, let us define the auxiliary matrix
\begin{equation}
\P\triangleq\begin{pmat}[{|}]
\mathbf{0}_{1\times (M-1)} & 1 \cr
\-
\I_{M-1} & \mathbf{0}_{(M-1)\times 1} \cr
\end{pmat}=
\begin{pmat}[{||}]
\P_1 & \cdots & \P_M
\cr\end{pmat}
\in \mathbb{C}^{M \times M},
\label{eq:pmatrixAppendix}
\end{equation}
where $\P_m$ denotes the $m$th column of $\P$, and such that the matrices $\M$, $\M^n$ and $(\M^n)^H$ appearing above can, referring to equation \eqref{COH:eq:permutation:matrix}, be respectively expressed as
\begin{subequations}
\begin{equation}
\M=\begin{pmat}[{|||}]\P_1 & \P_2 & \cdots & e^{ju_1}\P_M\cr\end{pmat},
\end{equation}
\begin{align}
\M^n\!&=\!\begin{pmat}[{|||||}]\!\P_1^n & \cdots\!\, & \P_{M-n}^n & e^{ju_1}\P_{M-n+1}^n&\cdots\!\,&e^{ju_1}\P_M^n\!\cr\end{pmat}\nonumber
\label{eq:M:n}
\\
&=\!\begin{pmat}[{|}]
\mathbf{0}_{n\times (M-n)} & \I_{n} e^{ju_1}\cr
\-
\I_{M-n} & \mathbf{0}_{(M-n)\times n} \cr
\end{pmat},\\
(\M^n)^H\!&=\!\begin{pmat}[{|||||}]e^{-ju_1}\P_{M-n+1}^n & \cdots\!\, & e^{-ju_1}\P_M^n &\P_1^n & \cdots\!\, & \P_{M-n}^n\cr\end{pmat}\!,\nonumber\\
&=\!\begin{pmat}[{|}]
\mathbf{0}_{(M-n)\times n} & \I_{M-n}\cr
\-
\I_{n} e^{-ju_1} & \mathbf{0}_{n\times (M-n)} \cr
\end{pmat}.
\end{align}
\end{subequations}

Using the expressions above, we obtain
\begin{subequations}
\begin{align}
\e_1^H\M^n\e_1
&=e_{n+1}^*e_1+\cdots+e_M^*e_{M-n}\nonumber\\
&\quad+e^{ju_1}\left(e_1^*e_{M-n+1} + \cdots + e_{n}^*e_M\right),
\label{COH:eq:e1H:Mn:e1}
\end{align}
\begin{align}
\e_1^H(\M^n)^H\e_1
&=e_{n+1}e_1^* + \cdots+e_M e_{M-n}^*\nonumber\\
&\quad+e^{-ju_1}\left(e_1e_{M-n+1}^*+ \cdots+e_{n}e_M^*\right)\nonumber\\
&=(\e_1^H\M^n\e_1)^*,
\end{align}
\label{COH:eq:M:MH}
\end{subequations}
which substituted into equation \eqref{COH:A:eq:coding:gain:AiAj} yields
\begin{align}
g(\e_1)
&=\!\!\!\!\!\!\min_{\substack{n\in \{1,\cdots,M-1\}\\s \in \mathcal{S}}}\!
\left(2-s\, \e_1^H\M^{n}\e_1+s^* (\e_1^H\M^{n}\e_1)^*\right)^\frac{1}{N}\nonumber\\
&=\!\!\!\!\!\!\min_{\substack{n\in \{1,\cdots,M-1\}\\s \in \mathcal{S}}}\!
\left( 2-s\,\e_1^H\M^{n}\e_1+(s\,\e_1^H\M^{n}\e_1)^*\right)^\frac{1}{N}\nonumber\\
&=\!\!\!\!\!\!\min_{\substack{n\in \{1,\cdots,M-1\}\\s \in \mathcal{S}}}\!
\left(2\,(1-\Re\{s\,\e_1^H\M^{n}\e_1\})\right)^\frac{1}{N}\!\!\!\!,\;\text{for } p < q,
\label{COH:A:eq:G2}
\end{align}
where $\Re\{\cdot\}$ denotes the real part of a complex number.

Finally, notice that $\M^{M-n}$ can be expressed as
\begin{align}
&\M^{M-n}\!=\!\begin{pmat}[{|||||}]\P_{M-n+1}^n & \cdots\!\, & \P_M^n & e^{ju_1}\P_1^n & \cdots\!\, & e^{ju_1}\P_{M-n}^n\cr\end{pmat} \nonumber\\
&=e^{ju_1}\!\!\begin{pmat}[{|||||}]\!e^{-ju_1}\P_{M-n+1}^n & \cdots\!\, & e^{-ju_1}\P_M^n & \P_1^n & \cdots\!\, & \P_{M-n}^n\cr\end{pmat}\nonumber\\
&=e^{ju_1}(\M^n)^H,
\label{COH:M:Mn}
\end{align}
which in turn implies the relations
\begin{align}
\Re\{s\, \e_1^H \M^{M-n}\e_1\}& = \Re\{s\, e^{ju_1}\, \e_1^H (\M^n)^H\e_1\}\\
&=\Re\{ (s\, \e_1^H\M^n\e_1)^*\}=\Re\{ s\, \e_1^H\M^n\e_1\},\nonumber
\end{align}
from which it is found that it is sufficient to consider $n\in\{1,\cdots,M/2\}$ in equation \eqref{COH:A:eq:G2}, as in equation \eqref{COH:propose:G2}, concluding the proof. \hfill $\square$

\section{Coding Gain of Sparse Projection Matrices}
\label{app:gain_sparse_e1}

We seek to prove that the coding gain under a dense sparse projection vector $\e_1$ designed via equation \eqref{COH:expand:basis} is the same as that under the dense vector obtained by solving problem \eqref{eq:minfe1}.
To this end, let us first rewrite equation \eqref{COH:propose:fanc} as
\begin{align}
f\left(\e_1(M,N_b)\right)=\sum_{n=1}^{M/2} F\left(\e_1(M,N_b),n\right),
\label{eq:rewrite:prop:func}
\end{align}
with
\begin{align}
F\left(\e_1(M,N_b),n\right)\triangleq|\e_1^H(M,N_b) \M^n \e_1(M,N_b)|,
\label{eq:Fdef}
\end{align}
where we remark that $\e_1^T(M,N_b)$ designed via equation \eqref{COH:expand:basis} is explicitly expressed by
\begin{equation}
\e_1^T(M,N_b)\! =\!\!\begin{pmat}[{||||||}]\! e_1\!\!\, & \mathbf{0}_{1 \times (\beta-1)} & e_2 & \mathbf{0}_{1 \times (\beta-1)} & \cdots\,\! & e_{N_b} & \mathbf{0}_{1 \times (\beta-1)}\!\!\cr\end{pmat}\!,
\label{eq:e1:T}
\end{equation}
where $\beta\triangleq(M/N_b)$ is a natural number.

Next, define $\gamma \triangleq n/ \beta$ and let the remainder of the division of $n$ by $\beta$ be denoted by $\mathrm{mod}(n,\beta)$, such that for all values of $n$ with $\mathrm{mod}(n,\beta)=0$, $\gamma\in\{1,\cdots,N_b/2\}$.
For these cases, using equations \eqref{eq:M:n} and \eqref{eq:e1:T}, the term $\e_1^H(M,N_b)\M^n$ in equation \eqref{eq:Fdef} can be expressed as
\begin{align}
\e_1^H(M,N_b)\M^n \!=\!&
\begin{pmat}[{|||||}. e^*_{\gamma+1} & \mathbf{0}_{1 \times (\beta-1)} & \cdots\!\, & e^*_{N_b} & \mathbf{0}_{1 \times (\beta-1)} \cr\end{pmat}& \nonumber \\
&\;\,\begin{pmat}.{||||}]
    e^{ju_1}e^*_1  & \mathbf{0}_{1 \times (\beta-1)} & \cdots\,\! & e^{ju_1}e^*_\gamma & \mathbf{0}_{1 \times (\beta-1)}\! \cr\end{pmat}\!\!\,.
\label{eq:e1:H:M:H}
\end{align}

Combining equations \eqref{eq:e1:T} and \eqref{eq:e1:H:M:H}, we obtain
\begin{align}
\e_1^H(M,N_b)\M^n\e_1(M,N_b)
&=e_{\gamma+1}^*e_1+\cdots+e_{N_b}^*e_{N_b-\gamma}\nonumber\\
&+e^{ju_1}\left(e_1^*e_{N_b-\gamma+1} + \cdots + e_{\gamma}^*e_{N_b}\right)\!\!\,.
\label{eq:new:e1H:Mn:e1}
\end{align}

Comparing the latter expression with equation \eqref{COH:eq:e1H:Mn:e1}, it is noticeable that $\gamma$ and $N_b$ in equation \eqref{eq:new:e1H:Mn:e1} play the same roles as $n$ and $M$ in equation \eqref{COH:eq:e1H:Mn:e1}, such that the norm of the quantity in equation \eqref{eq:new:e1H:Mn:e1} can be expressed using the notation introduced in equation \eqref{eq:Fdef}, $i.e.$ $|\e_1^H(M,N_b)\M^n\e_1(M,N_b)| = F(\e_1(N_b, N_b),\gamma)$.
But using equation \eqref{eq:Fdef} it then follows that
\begin{align}
F\left(\e_1(M,N_b),n\right) = F(\e_1(N_b, N_b),\gamma).
\label{eq:FdefNew}
\end{align}

In turn, for the values of $n$ such that $\mathrm{mod}(n,\beta)\neq0$, it is obvious that $\e_1^H(M,N_b)\M^n\e_1(M,N_b)=0$.
In other words, the terms of the sum in equation \eqref{eq:rewrite:prop:func}
are either $0$ or identical to the terms of the sum in equation \eqref{COH:propose:fanc}, $i.e.$
\begin{align}
f(\e_1(M,N_b)) = f(\e_1),
\label{eq:of:objective:func}
\end{align}
which implicates that the vector $\e_1(M,N_b)$ designed via equation \eqref{COH:expand:basis} achieves the same coding gain as that of $\e_1$, since it minimizes the same objective function in problem \eqref{eq:minfe1}, concluding proof. \hfill $\square$

\footnotesize{
	\bibliographystyle{IEEEtranURLandMonthDiactivated}
	\bibliography{main}

\begin{thebibliography}{10}
\providecommand{\url}[1]{#1}
\csname url@samestyle\endcsname
\providecommand{\newblock}{\relax}
\providecommand{\bibinfo}[2]{#2}
\providecommand{\BIBentrySTDinterwordspacing}{\spaceskip=0pt\relax}
\providecommand{\BIBentryALTinterwordstretchfactor}{4}
\providecommand{\BIBentryALTinterwordspacing}{\spaceskip=\fontdimen2\font plus
\BIBentryALTinterwordstretchfactor\fontdimen3\font minus
  \fontdimen4\font\relax}
\providecommand{\BIBforeignlanguage}[2]{{%
\expandafter\ifx\csname l@#1\endcsname\relax
\typeout{** WARNING: IEEEtran.bst: No hyphenation pattern has been}%
\typeout{** loaded for the language `#1'. Using the pattern for}%
\typeout{** the default language instead.}%
\else
\language=\csname l@#1\endcsname
\fi
#2}}
\providecommand{\BIBdecl}{\relax}
\BIBdecl

\bibitem{katsuki_optimal_2021}
Y.~Katsuki and N.~Ishikawa, ``Optimal but low-complexity optimization method
  for nonsquare differential massive {MIMO},'' in \emph{{IEEE} {Vehicular}
  {Technology} {Conference}}, virtual conference, Sep. 2021.

\bibitem{rivest_method_1978}
\BIBentryALTinterwordspacing
R.~L. Rivest, A.~Shamir, and L.~Adleman, ``\BIBforeignlanguage{en}{A method for
  obtaining digital signatures and public-key cryptosystems},''
  \emph{\BIBforeignlanguage{en}{Communications of the ACM}}, vol.~21, no.~2,
  pp. 120--126, 1978.
\BIBentrySTDinterwordspacing

\bibitem{shor_algorithms_1994}
\BIBentryALTinterwordspacing
P.~Shor, ``\BIBforeignlanguage{en}{Algorithms for quantum computation: discrete
  logarithms and factoring},'' in \emph{\BIBforeignlanguage{en}{Proceedings
  35th {Annual} {Symposium} on {Foundations} of {Computer} {Science}}}.\hskip
  1em plus 0.5em minus 0.4em\relax Santa Fe, NM, USA: IEEE Comput. Soc. Press,
  1994, pp. 124--134.
\BIBentrySTDinterwordspacing

\bibitem{fujii_noise_2016}
\BIBentryALTinterwordspacing
K.~Fujii, ``\BIBforeignlanguage{en}{Noise threshold of quantum supremacy},''
  \emph{\BIBforeignlanguage{en}{arXiv:1610.03632 [quant-ph]}}, 2016, arXiv:
  1610.03632.
\BIBentrySTDinterwordspacing

\bibitem{amico_experimental_2019}
\BIBentryALTinterwordspacing
M.~Amico, Z.~H. Saleem, and M.~Kumph, ``\BIBforeignlanguage{en}{Experimental
  study of {Shor}'s factoring algorithm using the {IBM} {Q} {Experience}},''
  \emph{\BIBforeignlanguage{en}{Physical Review A}}, vol. 100, no.~1, p.
  012305, 2019.
\BIBentrySTDinterwordspacing

\bibitem{BallNature2021}
P.~Ball, ``First quantum computer to pack 100 qubits enters crowded race,''
  \emph{Nature}, vol. 599, no. 542, 2021.

\bibitem{dean_physical-layer_2017}
\BIBentryALTinterwordspacing
T.~R. Dean and A.~J. Goldsmith, ``\BIBforeignlanguage{en}{Physical-layer
  cryptography through massive {MIMO}},'' \emph{\BIBforeignlanguage{en}{IEEE
  Transactions on Information Theory}}, vol.~63, no.~8, pp. 5419--5436, 2017.
\BIBentrySTDinterwordspacing

\bibitem{allen_secure_2014}
\BIBentryALTinterwordspacing
T.~Allen, J.~Cheng, and N.~Al-Dhahir, ``\BIBforeignlanguage{en}{Secure
  space-time block coding without transmitter {CSI}},''
  \emph{\BIBforeignlanguage{en}{IEEE Wireless Communications Letters}}, vol.~3,
  no.~6, pp. 573--576, 2014.
\BIBentrySTDinterwordspacing

\bibitem{althunibat_physical-layer_2017}
\BIBentryALTinterwordspacing
S.~Althunibat, V.~Sucasas, and J.~Rodriguez, ``\BIBforeignlanguage{en}{A
  physical-layer security scheme by phase-based adaptive modulation},''
  \emph{\BIBforeignlanguage{en}{IEEE Transactions on Vehicular Technology}},
  vol.~66, no.~11, pp. 9931--9942, 2017.
\BIBentrySTDinterwordspacing

\bibitem{okamoto_chaos_2012}
\BIBentryALTinterwordspacing
E.~Okamoto, ``\BIBforeignlanguage{en}{A chaos {MIMO} transmission scheme for
  channel coding and physical-layer security},''
  \emph{\BIBforeignlanguage{en}{IEICE Transactions on Communications}}, vol.
  E95.B, no.~4, pp. 1384--1392, 2012.
\BIBentrySTDinterwordspacing

\bibitem{ishikawa_articially_2021}
N.~Ishikawa, J.~M. Hamamreh, E.~Okamoto, C.~Xu, and L.~Xiao,
  ``\BIBforeignlanguage{en}{Artificially time-varying differential {MIMO} for
  achieving practical physical layer security},''
  \emph{\BIBforeignlanguage{en}{IEEE Open Journal of the Communications
  Society}}, vol.~2, pp. 1--15, 2021.

\bibitem{huang_fast_2013}
\BIBentryALTinterwordspacing
P.~Huang and X.~Wang, ``\BIBforeignlanguage{en}{Fast secret key generation in
  static wireless networks: {A} virtual channel approach},'' in
  \emph{\BIBforeignlanguage{en}{2013 {Proceedings} {IEEE} {INFOCOM}}}.\hskip
  1em plus 0.5em minus 0.4em\relax Turin, Italy: IEEE, Apr. 2013, pp.
  2292--2300.
\BIBentrySTDinterwordspacing

\bibitem{zeng_physical_2015}
\BIBentryALTinterwordspacing
K.~Zeng, ``\BIBforeignlanguage{en}{Physical layer key generation in wireless
  networks: challenges and opportunities},'' \emph{\BIBforeignlanguage{en}{IEEE
  Communications Magazine}}, vol.~53, no.~6, pp. 33--39, 2015.
\BIBentrySTDinterwordspacing

\bibitem{zhang_design_2017}
\BIBentryALTinterwordspacing
J.~Zhang, A.~Marshall, R.~Woods, and T.~Q. Duong,
  ``\BIBforeignlanguage{en}{Design of an {OFDM} physical layer encryption
  scheme},'' \emph{\BIBforeignlanguage{en}{IEEE Transactions on Vehicular
  Technology}}, vol.~66, no.~3, pp. 2114--2127, 2017.
\BIBentrySTDinterwordspacing

\bibitem{hamamreh_ofdm-subcarrier_2017}
\BIBentryALTinterwordspacing
J.~M. Hamamreh, E.~Basar, and H.~Arslan,
  ``\BIBforeignlanguage{en}{{OFDM}-subcarrier index selection for enhancing
  security and reliability of {5G} {URLLC} services},''
  \emph{\BIBforeignlanguage{en}{IEEE Access}}, vol.~5, pp. 25\,863--25\,875,
  2017.
\BIBentrySTDinterwordspacing

\bibitem{hughes_differential_2000}
B.~L. Hughes, ``\BIBforeignlanguage{en}{Differential space-time modulation},''
  \emph{\BIBforeignlanguage{en}{IEEE Transactions on Information Theory}},
  vol.~46, no.~7, p.~12, 2000.

\bibitem{ishikawa_rectangular_2017}
\BIBentryALTinterwordspacing
N.~Ishikawa and S.~Sugiura, ``\BIBforeignlanguage{en}{Rectangular differential
  spatial modulation for open-loop noncoherent massive-{MIMO} downlink},''
  \emph{\BIBforeignlanguage{en}{IEEE Transactions on Wireless Communications}},
  vol.~16, no.~3, pp. 1908--1920, 2017.
\BIBentrySTDinterwordspacing

\bibitem{ishikawa_differential_2018}
\BIBentryALTinterwordspacing
N.~Ishikawa, R.~Rajashekar, C.~Xu, S.~Sugiura, and L.~Hanzo,
  ``\BIBforeignlanguage{en}{Differential space-time coding dispensing with
  channel estimation approaches the performance of its coherent counterpart in
  the open-loop massive {MIMO}-{OFDM} downlink},''
  \emph{\BIBforeignlanguage{en}{IEEE Transactions on Communications}}, vol.~66,
  no.~12, pp. 6190--6204, 2018.
\BIBentrySTDinterwordspacing

\bibitem{ishikawa_differential-detection_2019}
\BIBentryALTinterwordspacing
N.~Ishikawa, R.~Rajashekar, C.~Xu, M.~El-Hajjar, S.~Sugiura, L.-L. Yang, and
  L.~Hanzo, ``\BIBforeignlanguage{en}{Differential-detection aided large-scale
  generalized spatial modulation is capable of operating in high-mobility
  millimeter-wave channels},'' \emph{\BIBforeignlanguage{en}{IEEE Journal of
  Selected Topics in Signal Processing}}, vol.~13, no.~6, pp. 1360--1374, 2019.
\BIBentrySTDinterwordspacing

\bibitem{xiao_differentially-encoded_2020}
\BIBentryALTinterwordspacing
L.~Xiao, P.~Xiao, H.~Ruan, N.~Ishikawa, L.~Lu, Y.~Xiao, and L.~Hanzo,
  ``\BIBforeignlanguage{en}{Differentially-encoded rectangular spatial
  modulation approaches the performance of its coherent counterpart},''
  \emph{\BIBforeignlanguage{en}{IEEE Transactions on Communications}}, vol.~68,
  no.~12, pp. 7593--7607, 2020.
\BIBentrySTDinterwordspacing

\bibitem{hochwald_differential_2000}
\BIBentryALTinterwordspacing
B.~Hochwald and W.~Sweldens, ``\BIBforeignlanguage{en}{Differential unitary
  space-time modulation},'' \emph{\BIBforeignlanguage{en}{IEEE Transactions on
  Communications}}, vol.~48, no.~12, pp. 2041--2052, 2000.
\BIBentrySTDinterwordspacing

\bibitem{chen-hu_non-coherent_2020}
\BIBentryALTinterwordspacing
K.~Chen-Hu, Y.~Liu, and A.~G. Armada, ``\BIBforeignlanguage{en}{Non-coherent
  massive {MIMO}-{OFDM} down-link based on differential modulation},''
  \emph{\BIBforeignlanguage{en}{IEEE Transactions on Vehicular Technology}},
  vol.~69, no.~10, pp. 11\,281--11\,294, 2020.
\BIBentrySTDinterwordspacing

\bibitem{xie_non-coherent_2020}
\BIBentryALTinterwordspacing
H.~Xie, W.~Xu, H.~Q. Ngo, and B.~Li, ``\BIBforeignlanguage{en}{Non-coherent
  massive {MIMO} systems: {A} constellation design approach},''
  \emph{\BIBforeignlanguage{en}{IEEE Transactions on Wireless Communications}},
  vol.~19, no.~6, pp. 3812--3825, 2020.
\BIBentrySTDinterwordspacing

\bibitem{rajashekar_algebraic_2017}
\BIBentryALTinterwordspacing
R.~Rajashekar, C.~Xu, N.~Ishikawa, S.~Sugiura, K.~V.~S. Hari, and L.~Hanzo,
  ``\BIBforeignlanguage{en}{Algebraic differential spatial modulation is
  capable of approaching the performance of its coherent counterpart},''
  \emph{\BIBforeignlanguage{en}{IEEE Transactions on Communications}}, pp.
  4260--4273, 2017.
\BIBentrySTDinterwordspacing

\bibitem{tarokh_differential_2000}
V.~Tarokh and H.~Jafarkhani, ``\BIBforeignlanguage{en}{A differential detection
  scheme for transmit diversity},'' \emph{\BIBforeignlanguage{en}{IEEE Journal
  on Selected Areas in Communications}}, vol.~18, no.~7, pp. 1169--1174, 2000.

\bibitem{hanzo_near-capacity_2009}
\BIBentryALTinterwordspacing
L.~Hanzo, O.~Alamri, M.~El-Hajjar, and N.~Wu,
  \emph{\BIBforeignlanguage{en}{Near-capacity multi-functional {MIMO}
  systems}}.\hskip 1em plus 0.5em minus 0.4em\relax Chichester, UK: John Wiley
  \& Sons, Ltd, May 2009.
\BIBentrySTDinterwordspacing

\bibitem{kraft_software_1988}
D.~Kraft, ``A software package for sequential quadratic programming,'' 1988.

\bibitem{buhmann_michael_2019}
\BIBentryALTinterwordspacing
M.~Buhmann, ``\BIBforeignlanguage{en}{Michael {J}.{D}. {Powell}'s work in
  approximation theory and optimisation},''
  \emph{\BIBforeignlanguage{en}{Journal of Approximation Theory}}, vol. 238,
  pp. 3--25, 2019.
\BIBentrySTDinterwordspacing

\bibitem{chhabra_design_2020}
\BIBentryALTinterwordspacing
S.~Chhabra, V.~Dhanwani, V.~K. Dhaka, and K.~Lata,
  ``\BIBforeignlanguage{en}{Design and analysis of secure one-way functions for
  the protection of symmetric key cryptosystems},'' in
  \emph{\BIBforeignlanguage{en}{2020 24th {International} {Symposium} on {VLSI}
  {Design} and {Test} ({VDAT})}}.\hskip 1em plus 0.5em minus 0.4em\relax
  Bhubaneswar, India: IEEE, Jul. 2020, pp. 1--6.
\BIBentrySTDinterwordspacing

\bibitem{wang_secrecy_2015}
\BIBentryALTinterwordspacing
L.~Wang, S.~Bashar, Y.~Wei, and R.~Li, ``\BIBforeignlanguage{en}{Secrecy
  enhancement analysis against unknown eavesdropping in spatial modulation},''
  \emph{\BIBforeignlanguage{en}{IEEE Communications Letters}}, vol.~19, no.~8,
  pp. 1351--1354, 2015.
\BIBentrySTDinterwordspacing

\bibitem{rezaei_aghdam_overview_2019}
\BIBentryALTinterwordspacing
S.~Rezaei~Aghdam, A.~Nooraiepour, and T.~M. Duman, ``\BIBforeignlanguage{en}{An
  overview of physical layer security with finite-alphabet signaling},''
  \emph{\BIBforeignlanguage{en}{IEEE Communications Surveys \& Tutorials}},
  vol.~21, no.~2, pp. 1829--1850, 2019.
\BIBentrySTDinterwordspacing

\bibitem{shu_two_2018}
\BIBentryALTinterwordspacing
F.~Shu, Z.~Wang, R.~Chen, Y.~Wu, and J.~Wang, ``\BIBforeignlanguage{en}{Two
  high-performance schemes of transmit antenna selection for secure spatial
  modulation},'' \emph{\BIBforeignlanguage{en}{IEEE Transactions on Vehicular
  Technology}}, vol.~67, no.~9, pp. 8969--8973, 2018.
\BIBentrySTDinterwordspacing

\bibitem{jiang_secrecy-enhancing_2018}
\BIBentryALTinterwordspacing
X.-Q. Jiang, M.~Wen, H.~Hai, J.~Li, and S.~Kim,
  ``\BIBforeignlanguage{en}{Secrecy-enhancing scheme for spatial modulation},''
  \emph{\BIBforeignlanguage{en}{IEEE Communications Letters}}, vol.~22, no.~3,
  pp. 550--553, 2018.
\BIBentrySTDinterwordspacing

\bibitem{soon_xin_ng_mimo_2006}
\BIBentryALTinterwordspacing
{Soon Xin Ng} and L.~Hanzo, ``\BIBforeignlanguage{en}{On the {MIMO} channel
  capacity of multidimensional signal sets},''
  \emph{\BIBforeignlanguage{en}{IEEE Transactions on Vehicular Technology}},
  vol.~55, no.~2, pp. 528--536, 2006.
\BIBentrySTDinterwordspacing

\bibitem{yilmaz_relationships_2020}
\BIBentryALTinterwordspacing
F.~Yilmaz, ``\BIBforeignlanguage{en}{On the relationships between average
  channel capacity, average bit error rate, outage probability, and outage
  capacity over additive white gaussian noise channels},''
  \emph{\BIBforeignlanguage{en}{IEEE Transactions on Communications}}, vol.~68,
  no.~5, pp. 2763--2776, 2020.
\BIBentrySTDinterwordspacing

\end{thebibliography}
}

\renewenvironment{IEEEbiography}[1]
{\IEEEbiographynophoto{#1}}
{\endIEEEbiographynophoto}

\begin{IEEEbiography}{Yuma~Katsuki}
(Member, IEEE) received the B.E. and M.E. degrees from Yokohama National University, Kanagawa, Japan, in 2021 and 2023, respectively. He is currently a researcher at NEC Corporation, Kanagawa, Japan.
\end{IEEEbiography}
\begin{IEEEbiography}{Giuseppe Thadeu Freitas de Abreu}(Senior Member, IEEE) received the B.Eng. degree in Electrical Engineering and a specialization (\emph{Latu Sensu}) degree in Telecommunications Engineering from the Universidade Federal da Bahia (UFBa), Salvador, Bahia, Brazil, in 1996 and 1997, respectively; and the M.~Eng. and D.~Eng. degrees in Physics, Electrical and Computer Engineering from the Yokohama National University, Japan, in March 2001, and March 2004, respectively, being the recipient of the Uenohara Award by Tokyo University in 2000 for his Master's Thesis work. He was a Post-doctoral Fellow and later Adjunct Professor (Docent) on Statistical Signal Processing and Communications Theory at the Department of Electrical and Information Engineering, University of Oulu, Finland from 2004 to 2006 and from 2006 to 2011, respectively. Since 2011 he is a Professor of Electrical Engineering at Jacobs University Bremen, renamed Constructor University in 2023. From April 2015 to August 2018 he also simultaneously held a full professorship at the Department of Computer and Electrical Engineering of Ritsumeikan University, Japan. His research interest span a wide range of topics within communications and signal processing, including communications theory, estimation theory, statistical modeling, wireless localization, cognitive radio, wireless security, MIMO systems, ultrawideband and millimeter wave communications, full-duplex and cognitive radio, compressive sensing, energy harvesting networks, random networks, connected vehicles networks, joint communications and sensing, and many others. He was the co-recipient of best paper awards at several international conferences, and was awarded JSPS, Heiwa Nakajima and NICT Fellowships (twice) in 2010, 2013, 2015 and 2018, respectively. Prof. Abreu served as an Associate Editor of the Transactions on Wireless Communications from 2009 to 2014, and of the Transactions on Communications from 2014 to 2017. He was an Executive Editor of the Transactions on Wireless Communications from 2018 to 2021 and since 2022, is serving as Editor to the IEEE Signal Processing Letters and the IEEE Communications Letters.
\end{IEEEbiography}
\begin{IEEEbiography}{Koji~Ishibashi}
(Senior Member, IEEE) received the B.E. and M.E. degrees in engineering from the University of Electro-Communications, Tokyo, Japan in 2002 and 2004, respectively, and the Ph.D. degree in engineering from Yokohama National University, Yokohama, Japan in 2007. From 2007 to 2012, he was an assistant professor in the Department of Electrical and Electronic Engineering, Shizuoka University, Hamamatsu, Japan. Since April 2012, he has been with the Communication Research Center (AWCC), the Advanced Wireless and University of Electro-Communications, where he is currently a professor. From 2010 to 2012, he was a visiting scholar at the School of Engineering and Applied Sciences at Harvard University, Cambridge, MA, USA. His current research interests include grant-free access, cell-free architecture, millimeter-wave communications, energy harvesting communications, wireless power transfer, channel codes, signal processing, and information theory.
\end{IEEEbiography}
\begin{IEEEbiography}{Naoki~Ishikawa}
(Senior Member, IEEE) received the B.E., M.E., and Ph.D. degrees from the Tokyo University of Agriculture and Technology, Tokyo, Japan, in 2014, 2015, and 2017, respectively. In 2015, he was an Academic Visitor with the School of Electronics and Computer Science, University of Southampton, U.K. From 2016 to 2017, he was a Research Fellow with the Japan Society for the Promotion of Science. From 2017 to 2020, he was an Assistant Professor with the Graduate School of Information Sciences, Hiroshima City University, Japan. He is currently an Associate Professor with the Faculty of Engineering, Yokohama National University, Kanagawa, Japan. His research interests include massive MIMO, physical layer security, and quantum speedup for wireless communications. He was certified as an Exemplary Reviewer of \textsc{IEEE Transactions on Communications} in 2017 and 2021.
\end{IEEEbiography}

\end{document}